\documentclass[11pt,showpacs,preprintnumbers,superscriptaddress,amsmath,amssymb,nofootinbib]{revtex4}
\usepackage{graphicx}
\usepackage{dcolumn}
\usepackage{bm}
\usepackage{amssymb}
\usepackage{amsmath}
\usepackage{epsfig}    
\usepackage{color}
\usepackage{slashed}
\usepackage{hhline}
\usepackage{mathrsfs}
\def\be{\begin{equation}}
\def\ee{\end{equation}}
\def\Mat3#1#2#3#4#5#6#7#8#9{
  \left(
  \begin{array}{ccc}
    #1 & #2 & #3 \\
    #4 & #5 & #6 \\
    #7 & #8 & #9 \\
  \end{array}
  \right) }
\newcommand{\bea}{\begin{eqnarray}}
\newcommand{\eea}{\end{eqnarray}}
\newcommand{\nn}{\nonumber}

\numberwithin{equation}{section}

\usepackage{slashed,color,graphicx}
\usepackage{hyperref}
\hypersetup{
  colorlinks = true,
  linkcolor = blue,
  citecolor = magenta
}

\begin{document}

%%%%%%%%%
\title{Muon $g-2$, dark matter, and neutrino mass explanations\\ in a modular $A_4$ symmetry }
\preprint{PKNU-NuHaTh-2020-07, KIAS-P20073, APCTP Pre2020-034}

\author{Parada~T.~P.~Hutauruk}
\email{phutauruk@gmail.com; phutauruk@pknu.ac.kr}
\affiliation{Department of Physics, Pukyong National University (PKNU), Busan 48513, Korea}
%

%\definechangesauthor[name=DW, color=green]{DW}
\author{Dong Woo Kang}
\email{dongwookang@kias.re.kr}
\affiliation{School of Physics, KIAS, Seoul 02455, Korea}

\author{Jongkuk Kim}
\email{jkkim@kias.re.kr}
\affiliation{School of Physics, KIAS, Seoul 02455, Korea}

\author{Hiroshi Okada}
\email{hiroshi.okada@apctp.org}
\affiliation{Asia Pacific Center for Theoretical Physics (APCTP) - Headquarters San 31, Hyoja-dong,
  Nam-gu, Pohang 790-784, Korea}
\affiliation{Department of Physics, Pohang University of Science and Technology, Pohang 37673, Republic of Korea}

\date{\today}

\begin{abstract}  
We study a successful model to explain the muon anomalous magnetic moment originating from Yukawa-type interactions {in a supersymmetric theory}. Thanks to a modular $A_4$ flavor symmetry, any lepton flavor violations that spoil the model are forbidden. 
We also investigate a predictive radiative seesaw model including a dark matter (DM) candidate.
At first, we construct the minimum model to satisfy the neutrino oscillation data and obtain several predictions such as Dirac CP and Majorana phases, the neutrino masses through $\chi^2$ analysis.
However, the minimum model would not provide our promising DM candidate.
{Thus, we minimally extend the model and find a good DM candidate.}
In the extended framework, we show the allowed regions to satisfy the muon anomalous magnetic moment and the observed relic density of dark matter in addition to predictions of the lepton sector.
\end{abstract}

\maketitle
\newpage

\section{Introduction}
Several flavor puzzles on muon anomalous magnetic moment (muon $g-2$ or $\Delta a_{\mu}$), neutrino masses, and mixings as well as phases are expected to be explained by theories beyond the Standard Model (SM). 
Recently, the muon anomalous magnetic moment {has attracted} more attention from particle physicists since the discrepancy between the experiment and the SM predictions was confirmed. 
To resolve the anomaly of $\Delta a_{\mu}$, one must introduce an extra charged fermion or at least a new boson that couples to muon. 
On the other hand, a radiative seesaw model is believed to be one of the elegant solutions to explain neutrino oscillation data as well as $\Delta a_{\mu}$ data at a low energy scale.
In Ref.~\cite{Ma:2006km,Tao:1996vb},  heavy Majorana fermions are introduced to generate a one-loop diagram of the active neutrino mass matrix.
The Large Hadron Collider (LHC) can be used to search for the footprint of the model. 
This implies that the small Yukawa couplings are unnecessary even if their extra masses are of the 1 TeV scale.
The implementation of the loop diagram is imposed based on a (residual) symmetry, however, it depends on the model construction.
This also leads us to explain DM candidate as well as $\Delta a_{\mu}$ \footnote{ You can also see $U(1)_{L_\mu-L_\tau}-$charged DM model to resolve both muon $g-2$ anomaly and the correct DM relic density with much wider DM mass range \cite{Baek:2022ozm}.}.
In such a scenario, we have to consider constraints of the lepton flavor violations (LFVs) such as $\mu\to e\gamma$, where it is rather difficult to constraint. 

Recently, attractive flavor symmetries are proposed in Ref.~\cite{deAdelhartToorop:2011re,Feruglio:2017spp}.
They have applied modular symmetry motivated by non-Abelian discrete flavor symmetries to quark and lepton sectors.
One remarkable advantage of applying this symmetry is the dimensionless couplings of the model can be transformed into non-trivial representations under those symmetries. We then do not need the scalar fields to obtain a predictive mass matrix.
Along with this idea, a vast reference has recently appeared in the literature, {\it e.g.}, $A_4$~\cite{Feruglio:2017spp, Criado:2018thu,Kobayashi:2018scp, Okada:2018yrn,Nomura:2019jxj, Okada:2019uoy, deAnda:2018ecu, Novichkov:2018yse, Nomura:2019yft, Okada:2019mjf,Ding:2019zxk, Nomura:2019lnr,Kobayashi:2019xvz,Asaka:2019vev,Zhang:2019ngf, Ding:2019gof,Kobayashi:2019gtp,Nomura:2019xsb, Wang:2019xbo,Okada:2020dmb,Okada:2020rjb, Behera:2020lpd, Behera:2020sfe, Nomura:2020opk, Nomura:2020cog, Asaka:2020tmo, Okada:2020ukr, Nagao:2020snm, Okada:2020brs,Kang:2022psa,Kim:2023jto},
%%%
$S_3$ \cite{Kobayashi:2018vbk, Kobayashi:2018wkl, Kobayashi:2019rzp, Okada:2019xqk, Mishra:2020gxg, Du:2020ylx},
%%%
$S_4$ \cite{Penedo:2018nmg, Novichkov:2018ovf, Kobayashi:2019mna, King:2019vhv, Okada:2019lzv, Criado:2019tzk,
Wang:2019ovr}, $A_5$~\cite{Novichkov:2018nkm, Ding:2019xna,Criado:2019tzk}, double covering of $A_5$~\cite{Wang:2020lxk, Yao:2020zml}, larger groups~\cite{Baur:2019kwi}, multiple modular symmetries~\cite{deMedeirosVarzielas:2019cyj}, and double covering of $A_4$~\cite{Liu:2019khw, Chen:2020udk}, $S_4$~\cite{Novichkov:2020eep, Liu:2020akv}, and the other types of groups \cite{Kikuchi:2020nxn} in which masses, mixing, and CP phases for the quark and/or lepton have been predicted~\footnote{For interest readers, we provide some literature reviews, which are useful to understand the non-Abelian group and its applications to flavor structure~\cite{Altarelli:2010gt, Ishimori:2010au, Ishimori:2012zz, Hernandez:2012ra, King:2013eh, King:2014nza, King:2017guk, Petcov:2017ggy}.}.
Moreover, a systematic approach to understanding the origin of CP transformations has been discussed in Ref.~\cite{Baur:2019iai}, 
and CP violation in models with modular symmetry was discussed in Refs.~\cite{Kobayashi:2019uyt,Novichkov:2019sqv}, 
and a possible correction from K\"ahler potential was discussed in Ref.~\cite{Chen:2019ewa}. Furthermore,
a systematic analysis of the fixed points (stabilizers) has been discussed in Ref.~\cite{deMedeirosVarzielas:2020kji}.
A very recent paper of Ref.~\cite{Ishiguro:2020tmo} finds a favorable fixed point $\tau=\omega$ among three fixed points, which are the fundamental domain of PSL$(2,Z)$, by systematically analyzing the stabilized moduli values in the possible configurations of flux compactifications as well as investigating the probabilities of moduli values.
% showing which moduli values are favorable from our moduli stabilization.

In this paper, we successfully introduce a term in our construction model to explain the muon anomalous magnetic moment without suffering from the LFVs using the modular $A_4$ symmetry {in a supersymmetric theory}. In addition, we construct two predictive neutrino mass models based on a radiative seesaw model. Even though we have obtained several predictions for the minimum model, the minimum model would not provide our promising DM candidate.
 Thus, we minimally extend the model and get our DM candidate.
 In the extended framework, we show the allowed regions to satisfy the muon anomalous magnetic moment and the observed relic density of dark matter in addition to predictions of the lepton sector.
 We have performed numerical analysis for the lepton sector, applying the $\chi^2$ fit at 2, 3, and 5 $\sigma$ confidence level. 
%We also show the allowed region for each of the C.L. and we focus on the region near the most favorable fixed point of $\tau=\omega$ as in the recent theoretical analysis of Ref.~\cite{Ishiguro:2020tmo}.

This paper is organized as follows.
In Sec.~\ref{model:setup}, we define our model construction and discuss various phenomenological objects under our model including heavier and active neutral neutrino fermion mass matrices, and LFVs.
In Sec.~\ref{Muon:DM}, we formulate
the muon anomalous magnetic moments and the relic density of the bosonic DM candidate.
%{we define our model construction where in the formula we consider the muon anomalous magnetic moments, neutral and active neutrino fermion mass matrices, LFVs, and the relic density of the bosonic/fermionic DM candidate.}
In Sec.~\ref{sec:NAl}, we present our numerical analysis and several predictions such as Dirac CP and Majorana phases, the neutrino masses through $\chi^2$ analysis for two scenarios. 
Then, we search for the preferred region to satisfy the muon $g-2$ and DM relic density selecting our favorite model in the neutrino analysis.
We conclude in Sec.~\ref{con}. In the Appendix, we review the modular group and we show how the multiplication rules work in $A_4$ symmetry.

%%%%%%%%%%%%%%%%%%%%%%%%%%%%%%%%%%%%%
% \subsection{Particle contents and the model Lagrangian}
\begin{widetext}
  \begin{center} 
    \begin{table}%[tbc]
      \begin{footnotesize}
        \begin{tabular}{|c||c|c|c|c|c|c|c|c|c|c|}\hline\hline
          & \multicolumn{9}{c|}{Leptons} \\\hline
            ~&~ $\hat L_{e}$ ~&~ $\hat L_{\mu}$ ~&~ $\hat L_\tau$ ~&~ $\hat {\bar e}$ ~&~ $\hat {\bar \mu}$ ~&~ $\hat {\bar \tau}$  ~&~ $\hat{E}$ ~&~ $\hat{\bar E}$ ~&~ $\hat {\bar N}$~
          \\\hline 
          %$SU(3)_C$  & $\bm{1}$  & $\bm{1}$  & $\bm{1}$   & $\bm{1}$  & $\bm{1} $  & $\bm{1}$ & $\bm{1}$  & $\bm{1} $    \\\hline 
          %%%
          $SU(2)_L$  & $\bm{2}$  & $\bm{2}$  & $\bm{2}$   & $\bm{1}$  & $\bm{1}$   & $\bm{1}$ & $\bm{1}$ & $\bm{1}$  & $\bm{1}$  \\\hline 
          %%%
          $U(1)_Y$  & $\frac{1}{2}$ & $\frac12$ & $\frac12$  & $-1$ &  $-1$  &  $-1$ & ${1}$& ${-1}$ &  $0$    \\\hline
          %%%
          $A_4$ & $1$  & $1'$ & $1''$ & $1$  & $1''$   & $1'$ & $1'$ & $1''$  & $3$   \\\hline
          %%%
          $-k$  & $0$  & $-2$ & $0$ & $0$ & $-2$ & $0$ & $-1$& $-3$ & $-k_N$   \\\hline
          %%%
          %$\mathbb{Z}_2$ & $+$   & $-$  & $+$ & $+$& $-$& $-$& $+$ & $+$  \\\hline\hline
        \end{tabular}
        \caption{Field contents of the matter superfields
          and their charge assignments under $SU(2)_L\otimes U(1)_Y\otimes A_4$,
          where
          $SU(3)_C$ singlet for all the SM leptons and new fields, and $-k$ is the number of modular weight and $k_N=1,3$.}
        \label{tab:1}
      \end{footnotesize}
    \end{table}
  \end{center}
\end{widetext}

\begin{table}[t]
\centering {\fontsize{10}{12}
\begin{tabular}{|c||c|c||c|c|c|c|  }\hline\hline
&\multicolumn{2}{c||}{VEV$\neq 0$} & \multicolumn{3}{c|}{Inert } \\\hline
    &~ $\hat H_1$  ~ &~ $\hat H_2$  ~   &~ $\hat \eta_1$ ~&~ $\hat \eta_2$ ~ &~ $\hat \chi$~      \\\hline
$SU(2)_L$ & $\bm{2}$  & $\bm{2}$ & $\bm{2}$   & $\bm{2}$  & $\bm{1}$       \\\hline 
$U(1)_Y$ & $\frac12$    & $-\frac12$& $\frac12$    & $-\frac12$  & $0$       \\\hline
 $A_4$  & $1$   & $1$ & $1$  & $1$ & $1$     \\\hline
$-k$ & $0$ & $0$  & $-1$    & $-3$ & $-3$    \\\hline
\end{tabular}%
} 
\caption{Field contents of superfields
and their charge assignments under $SU(2)_L\otimes U(1)_Y\otimes A_4$, where $SU(3)_C$ singlet for all bosons and $-k$ is the number of modular weight. }
\label{tab:2}
\end{table}

%\subsection{Lagrangian}

 \section{Model setup} \label{model:setup}
In this Section, we explain our model construction
by introducing new fields and assigning charges under the symmetries of  $SU(2)_L\otimes U(1)_Y\otimes A_4$ into the lepton and Higgs sectors, where upper "hat" indices of fields represent superfields and $SU(3)_C$ singlet for all bosons and $-k$ is the number of modular weight.
 For the fermion sector, we add one vector-like matter superfield ($\hat E, \hat{\bar E}$) including a singly-charged heavy lepton $E, \bar E$,  and three neutral matter superfields $\hat{\bar N}$ including Majorana fermions $\bar N\equiv [\bar N_{e},\bar N_{\mu},\bar N_{\tau}]^T$. 
The $\hat E, \hat{\bar E}$ has $1'_{-1},1''_{-3}$ under $A_4$ and modular weight, and the $\hat{\bar N}$ has $A_4$ triplet under $-k_{N}=-1,-3$. 
 %%%
The fermionic contents and their charged assignments are shown in Table~\ref{tab:1}.

For the boson sector, we introduce superfields $\hat \chi, \hat\eta_1,\hat\eta_2$ including an isospin singlet inert boson $\chi$, and doublet inert one $\eta_1,\eta_2$, where all have zero vacuum expectation values (VEVs). 
%where the charges under the SM gauge symmetry are the same as the SM Higgs.
The $\chi$ is denoted by $\chi=(\chi_R + i\chi_I)/\sqrt2$ that is $A_4$ trivial singlet with $-3$ modular weight.
%The $\chi$ plays a role in inducing the muon $g-2$ together with $E$ and $L'$.
%
The $\eta_1, \eta_2$ are respectively denoted by $\eta_1=[\eta^+_1,(\eta_{1R}+i\eta_{1I})/\sqrt2]^T$ and  $\eta_2=[(\eta_{2R}+i\eta_{2I})/\sqrt2, \eta^-_2]^T$ that are $A_4$ trivial singlets with $-1, -3$ modular weights.
%The $\eta$ plays a role in generating both the electron $g-2$ and the neutrino mass matrix together with $N_R$.
%
 Superfields $\hat H_1$ and $\hat H_2$ are requested by supersymmetry to cancel anomaly as usual and
we denote their bosonic parts are written by $H_1=[h_1^+,(v_1+h_1+i z_1)/\sqrt2]^T$ and $H_2=[(v_2+h_2+i z_2)/\sqrt2,h_2^-]^T$ with totally neutral charges under $A_4$ and $(-k)$, therefore the structure is exactly the same as the minimum supersymmetric theory. The SM VEV is defined by $v_H\equiv \sqrt{v_1^2+v_2^2}\equiv 246$ GeV. 
The bosonic field contents and their charge assignments are listed in Table~\ref{tab:2}.
The $\bar E, E, \chi, \eta_2$ play a key role in generating the muon $g-2$, while the $\bar N, \chi,\eta_1$ contribute to
the neutrino mass matrix at the one-loop level. 
%Notice here that only $E$ cannot explain sizable muon $g-2$ due to chiral suppression.
%
Under these symmetries, one writes the valid superpotential as follows:
\begin{align}
 \mathcal{W}_Y & = 
  y_e \hat {\bar e} \hat H_2\hat L_e  +y_\mu \hat {\bar \mu} \hat H_2\hat L_\mu  + y_\tau   \hat{\bar\tau} \hat H_2 \hat L_\tau 
+ h \hat L_\mu \hat\eta_2 \hat {\bar E}+ y_E \hat{\bar\mu} \hat E \hat\chi
{+ y'_E \hat{\bar\tau} \hat E \hat\chi}
 \nn\\
%+ h \overline{E_L} L'_R H^*  }
  %%%
  &+a_\eta [Y^{(1+k_N)}_3\otimes \hat {\bar N}] \hat L_e \hat \eta_1
  +b_\eta [Y^{(3+k_N)}_3\otimes \hat {\bar N}] \hat L_\mu \hat \eta_1
  +c_\eta [Y^{(1+k_N)}_3\otimes \hat {\bar N}] \hat L_\tau \hat \eta_1 
  +M_0 [Y^{(2k_N)}_3\otimes \hat {\bar N} \otimes \hat {\bar N}] 
    \nn\\
  & + M_E \hat{\bar E} \hat E  +\mu_H\hat H_1\hat H_2+\mu_\eta \hat \eta_1\hat \eta_2+\mu_\chi \hat \chi\hat \chi+a\hat H_2 \hat\eta_1\hat\chi
  +b\hat H_1 \hat\eta_2\hat\chi ,
  %\nn\\
%  &+a_\chi [(Y^{(2)}_3)^* \otimes \overline{ e_{R}}\otimes N_R^C] \chi^-
%  +b_\chi [(Y^{({\color{red}4})}_3)^*\otimes \overline{ \mu_{R}}\otimes N_R^C]  \chi^-
%  + c_\chi [(Y^{(2)}_3)^*\otimes \overline{ \tau_{R}} \otimes N_R^C] \chi^-
  %%%
\label{yukawa}
\end{align}
where the parentheses of $[\cdots]$ represent singlet under $A_4$ by applying its multiplication rules; Yukawa matrices have concrete structures as discussed later.
%{we neglect the term proportional to $a$ for simplicity that induces mixing between $N_R$ and neutral component of $L'$. Even with nonzero mixing, our result does not change much.}
%the charged-lepton mass matrix is diagonal due to the modular $A_4$ symmetry, and $a_\chi, b_\chi, c_\chi$ are the complex parameters.
%%%
%The Higgs potential is also given by~\cite{Ma:2006km}
Valid soft SUSY-breaking terms to construct the neutrino mass matrix and muon $g-2$ are found as follows: 
\begin{align}
-{\cal L}_{\rm soft} \sim& \mu_{BH}^2 H_1 H_2 +
 \mu_{B\eta}^2 \eta_1\eta_2+ \mu_{B\chi}^2 \chi\chi +
A_a  H_2\eta_1 \chi+ A_b H_2\eta_2 \chi\nn\\
%%%
&
+m^2_{H_1}|H_1|^2+m^2_{H_2}|H_2|^2
+m^2_{\eta_1}|\eta_1|^2+m^2_{\eta_2}|\eta_2|^2+m^2_{\chi}|\chi|^2
+ {\rm h.c.}.\nn \label{eq:pot}
\end{align}

\subsection{Neutral fermions}
The Yukawa mass matrix coming from $a_\eta,b_\eta,c_\eta$, denoted by {$\bar N Y_\eta L \eta_1
\sim   \bar N Y_\eta \nu(\eta_{1R}-i\eta_{1I})/\sqrt2$, are written as follows:
\begin{align}
&k_N=1,\quad Y_\eta^T  = a_\eta
\begin{pmatrix}
1 & 0 & 0 \\ 
 0 &  b_\eta/a_\eta &0 \\ 
0 & 0 &  c_\eta/a_\eta \\ 
\end{pmatrix}
%%%
\begin{pmatrix}
 y_1 & y_3 & y_2 \\ 
 y_1^{(4)} &  y_3^{(4)} & y_2^{(4)} \\ 
y_2 & y_1 & y_3 \\ 
\end{pmatrix}
\equiv a_\eta \tilde Y_\eta^T,\\
%%% %%%
&k_N=3,\quad Y_\eta^T  = a_\eta
\begin{pmatrix}
1 & 0 & 0 \\ 
 0 &  b_\eta/a_\eta &0 \\ 
0 & 0 &  c_\eta/a_\eta \\ 
\end{pmatrix}
%%%
\begin{pmatrix}
 y_1^{(4)} & y_3^{(4)} & y_2^{(4)} \\ 
 y_1^{(6)}+\epsilon y'^{(6)}_1 &  y_3^{(6)}+\epsilon y'^{(6)}_3 & y_2^{(6)}+\epsilon y'^{(6)}_2 \\ 
y_2^{(4)} & y_1^{(4)} & y_3^{(4)} \\ 
\end{pmatrix}
\equiv a_\eta \tilde Y_\eta^T,
\end{align}
}
where $Y^{(4)}_3\equiv  [y_1^{(4)} ,  y_2^{(4)} , y_3^{(4)}]^T$, $Y^{(6)}_3\equiv  [y^{(6)}_1 ,  y^{(6)}_2, y^{(6)}_3]^T$,
$Y'^{(6)}_3\equiv  [y'^{(6)}_1,  y'^{(6)}_2, y'^{(6)}_3]^T$ are given in the Appendix, and $a_\eta,b_\eta,c_\eta$ stand for the real parameters by the phase redefinition of fields, while $\epsilon$ is a complex free parameter.

The Majorana heavier neutrino mass matrix is given by
\begin{align}
&k_N=1,\quad
\mathcal{M}_N  = 
\frac{M_0}{3}
\begin{pmatrix}
2 y_1 &- y_3 & - y_2 \\ 
 -y_3 & 2 y_2 & - y_1 \\ 
- y_2 &- y_1 & 2 y_3 \\ 
\end{pmatrix}
\equiv
\frac{M_0}{3}\tilde {\mathcal{M}}_N
,\label{massmat}\\
%%% %%%
&k_N=3,\quad
\mathcal{M}_N  = 
\frac{M_0}{3}
{\begin{pmatrix}
2 (y^{(6)}_1 +r_1 y'^{(6)}_1) +r_2 &- y^{(6)}_3 -r_1 y'^{(6)}_3 & - y^{(6)}_2 -r_1 y'^{(6)}_2 \\ 
 -y^{(6)}_3-r_1 y'^{(6)}_3 & 2 (y^{(6)}_2+r_1 y'^{(6)}_2) & - y^{(6)}_1 - r_1y'^{(6)}_1 + r_2\\ 
- y^{(6)}_2 -r_1 y'^{(6)}_2 &- y^{(6)}_1 -r_1 y'^{(6)}_1 +r_2 & 2 (y^{(6)}_3 + r_1 y'^{(6)}_3) \\ 
\end{pmatrix}
}
\equiv
\frac{M_0}{3}\tilde {\mathcal{M}}_N
,\label{massmat}
\end{align}
where $M_0$ is taken to be real without loss of generality and $r_{1,2}$ are complex free parameters.

\begin{figure}[htbp]
  \includegraphics[width=85mm]{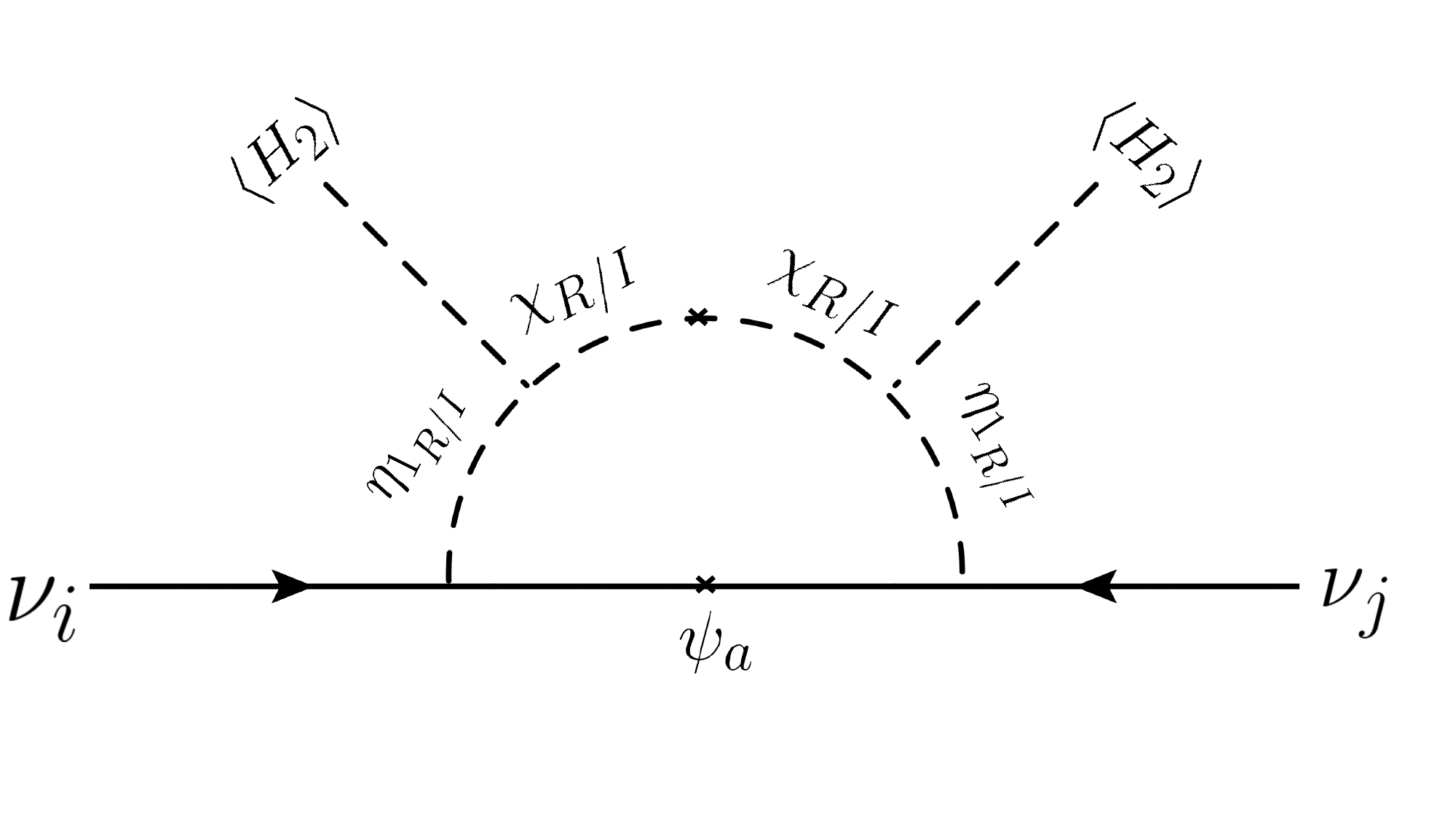}
  \caption{Feynman diagram for 1-loop neutrino mass generation.}
  \label{fig:neut-fm}
\end{figure}
%%%

The mass matrix $\tilde {\mathcal{M}}_N$ is then diagonalized by multiplying it with a
unitary matrix $V$, it then gives
\begin{align}
V^T \tilde {\mathcal{M}}_N V=\tilde {\mathcal{M}}_N^\text{diag} \equiv \text{diag}(\tilde M_{1},\tilde M_{2},\tilde M_{3}). 
\end{align}
Here, the mass eigenstate $\psi_R$ is defined by $N_{R_i}=\sum_{k=1,3}V_{i k} \psi_{R_k}$, and its mass eigenvalue is defined by $M_a=\frac{M_0}3\tilde M_{a}(a=1,2,3)$.
%%%%%%%%%%%%%%%%%%%
{The active neutrino mass matrix is generated via 1-loop as shown in Fig.~\ref{fig:neut-fm}.}
%%%%%%%%%%%%%%%%%%%
To induce the neutrino mass matrix into the Lagrangian,
we rewrite the Lagrangian in terms of the mass eigenstate as
\begin{align}
-{\cal L}_\nu&= \frac{a_\eta}{\sqrt2}\bar\psi_{a} \tilde F_{ai} \nu_{i} (\eta_{1R}+i \eta_{1I})+{\rm h.c.}+\frac{A_a v_2}{\sqrt2}(\eta_{1R}\chi_{R}-\eta_{1I}\chi_{I}) + \mu^2_{B_\chi}(\chi^2_R-\chi^2_I) 
,\label{eq:neut}
\end{align}
where $\tilde F\equiv V^T \tilde  Y_\eta$.
The mass matrix is given by 
\begin{align}
  (m_{\nu})_{ij}
&\simeq -\frac{a^2_\eta M_0}{12(4\pi)^2}\frac{A_a^2 v_2^2}{m^4_{\chi_R}}
%\left(1-\frac{m^2_{\chi_I}}{m^2_{\chi_R}}\right)
\left(1-r_{\chi_I}\right)
\sum_{a=1}^{3} \tilde F_{i\alpha}^T \tilde M_\alpha \tilde F_{\alpha j} F_{0},\\
\label{massmatrix2}
F_{0}&=\int[dx]_4 \frac{x_1}{(x_1 r_{\eta_1} + x_2 + x_3 r_{\chi_I}+x_4 r_{\psi_a})^3},
\end{align}
where we assume $m_{\eta_1}\equiv m_{\eta_{1R}}= m_{\eta_{1I}}$, $\int[dx]_4\equiv\int dx_1dx_2dx_3dx_4\delta(x_1+x_2+x_3+x_4-1)$, $r_f\equiv m^2_{f}/m^2_{\chi_R}$ ($f\equiv \eta_1,\chi_I,\psi_a$) $m_{\psi_a}\equiv M_a$, and we have applied mass insertion approximation to this loop calculation.~\footnote{It suggests that square masses of $\eta_{1R,1I}, \chi_{R,I}$ have to be greater than $A_a v_2/\sqrt2$.}
The neutrino mass matrix $m_\nu$ is diagonalized by using a unitary matrix $U_{\rm PMNS}$~\cite{Maki:1962mu}; $D_\nu\equiv U_{\rm PMNS}^T m_\nu U_{\rm PMNS}$.
Here, we define the dimensionless neutrino mass matrix as $m_\nu \equiv\kappa \tilde m_\nu$,
where the $\kappa\equiv -\frac{a^2_\eta M_0}{12(4\pi)^2}\frac{A_a^2 v_2^2}{m^4_{\chi_R}}
\left(1-r_{\chi_I}\right)$ does not depend on the flavor structure. The diagonalization in terms of the dimensionless form $\tilde D_\nu\equiv U_{\rm PMNS}^T \tilde m_\nu U_{\rm PMNS}$ is rewritten.
Thus, we fix $\kappa$ by using this formula
\begin{align}
({\rm NH}):\  \kappa^2= \frac{|\Delta m_{\rm atm}^2|}{\tilde D_{\nu_3}^2-\tilde D_{\nu_1}^2},
\quad
({\rm IH}):\  \kappa^2= \frac{|\Delta m_{\rm atm}^2|}{\tilde D_{\nu_2}^2-\tilde D_{\nu_3}^2},
 \end{align}
where $\tilde m_\nu$ is diagonalized by $V^\dag_\nu (\tilde m_\nu^\dag
\tilde m_\nu)V_\nu=(\tilde D_{\nu_1}^2,\tilde D_{\nu_2}^2,\tilde
D_{\nu_3}^2)$  and $\Delta m_{\rm atm}^2$ is the atmospheric neutrino
mass-squared difference. NH and IH stand for normal
and the inverted hierarchies, respectively. Subsequently, the solar neutrino mass-squared difference is described in terms of the $\kappa$ as follows:
\begin{align}
\Delta m_{\rm sol}^2= {\kappa^2}({\tilde D_{\nu_2}^2-\tilde D_{\nu_1}^2}).
 \end{align}
This should be within the range of the experimental value. Later, we will adopt NuFit 5.1~\cite{Gonzalez-Garcia:2021dve} to our numerical analysis. 
The neutrinoless double beta decay is also given by 
\begin{align}
\langle m_{ee}\rangle=\kappa|\tilde D_{\nu_1} \cos^2\theta_{12} \cos^2\theta_{13}+\tilde D_{\nu_2} \sin^2\theta_{12} \cos^2\theta_{13}e^{i\alpha_{2}}+\tilde D_{\nu_3} \sin^2\theta_{13}e^{i(\alpha_{3}-2\delta_{CP})}|.
\end{align}
This may be able to observe in the future experiments of KamLAND-Zen~\cite{KamLAND-Zen:2016pfg}.

%%%%%%%%%%%%%%%%%%%%%%%%%%%%%%%%%%%%%%%%%%%%%%%%%%
\subsection{Lepton flavor violations related to neutrino masses} \label{lfv-lu}
%%%%%%%%%%%%%%%%%%%%%%%%%%%%%%%%%%%%%%%%%%%%%%%%%%
Since the term that generates the neutrino mass matrix also arises lepton flavor violating processes at the one-loop level, we need to check whether these constraints are satisfied or not.
The valid Lagrangian to arise LFVs comes from the same term that generates the neutrino mass matrix and it is written in terms of the mass eigenstate as follows:
{\begin{align}
-{\cal L}_\nu&= {a_\eta}
\bar\psi_{a} \tilde F_{ai} \ell_{i} \eta^+_{1}
+{\rm h.c.},
\label{eq:lvs-g2}
\end{align}
%where we define {$G_e^a \equiv a_\chi  [y_1^* V^*_{1a}+ y_3^* V^*_{2a}+y_2^* V^*_{3a}]$, $G_\tau^a \equiv c_\chi  [y_1^* V^*_{2a}+ y_3^* V^*_{3a}+y_2^* V^*_{1a}]$}, and $a$ should be summed up over $1-3$.
The corresponding branching ratio is given at one-loop level as follows~\cite{Lindner:2016bgg, Baek:2016kud, Toma:2013zsa}
\begin{align}
&{\rm BR}(\ell_i\to\ell_j\gamma)
=
{a_\eta^4} \frac{48\pi^3\alpha_{\rm em} C_{ij}}{(4\pi)^4 G_F^2}
\left| \sum_{a=1}^3\tilde F^\dag_{ja}\tilde F_{ai} \Pi(M_a,m_{\eta^-})  \right|^2
\left(1+\frac{m_j^2}{m_i^2}\right),\\
%%%
%&A_{ij}=\frac{1}{(4\pi)^2} \sum_{a=1}^3 \left(F^\dag_{ja}F_{ai} II(M_a,m_{\eta^-}) { +(|\delta_{ie}G_i^a|^2+|\delta_{i\tau}G_i^a|^2) II(M_a,m_{\chi^-}) }\right),\\
%%%
&\Pi(m_1,m_2)\simeq 
\frac{m_2^6 -6 m_2^4 m_1^2 + 3 m_2^2 m_1^4 +2 m_1^6+6 m_2^2 m_1^4\ln\left[\frac{m^2_2}{m^2_1}\right]}{12(m_2^2-m_1^2)^4},
\label{eq:damu1}
\end{align}
}
where $i,j$ runs over $e,\mu, \tau$, the fine structure constant $\alpha_{\rm em} \simeq 1/128$, the Fermi constant $G_F \simeq 1.17\times 10^{-5}$ GeV$^{-2}$, and $(C_{21}, C_{31}, C_{32}) \simeq (1, 0.1784, 0.1736)$. ${\rm II}(m_1,m_2) $ is derived by assuming $m_{i,j} \ll M_a, m_{\eta^-}$, and notice ${\rm II}(m_1,m_2) = \frac{1}{24M_a^2}$ in the limit of $M_a=m_{\eta^-}$. 
The current experimental upper bounds at 90\% C.L. are~\cite{MEG:2016leq,MEG:2013oxv}
\begin{align}
{\rm BR}(\mu\to e\gamma) < 4.2\times10^{-13} ~,~
{\rm BR}(\tau\to e\gamma) < 3.3\times10^{-8} ~,~
{\rm BR}(\tau\to \mu\gamma) < 4.4\times10^{-8} ~.\label{eq:exp-lfvs}
\end{align}

\section{Muon anomalous magnetic dipole moment, LFVs, and dark matter} \label{Muon:DM}

\subsection{Muon anomalous magnetic dipole moment: muon $g-2$($\Delta a_\mu$)} \label{lfv-lu}
The muon anomalous magnetic dipole moment has been first reported by Brookhaven National Laboratory (BNL). 
They reported that the muon $g-2$ data has a discrepancy at the 3.3$\sigma$ level from the SM prediction. 
Recent experimental result~\cite{Hanneke:2008tm} of muon $g-2$ suggests the following value at $4.2\sigma$~\cite{Muong-2:2021ojo}:
\begin{align}
\Delta a_\mu = a_\mu^{\rm{EXP}}-a_\mu^{\rm{SM}} = (25.1\pm 5.9)\times 10^{-10}. \label{eq:yeg2}
\end{align}
%%%%%%%%%%%%%%%%%%%%%%%%%%%%%%%%%%%%%%%%%%%%%%%%%%

%Although 
Here we consider the above result, however, it is worth mentioning other results~\footnote{We thank the referee for pointing it out.}. If we take into account recent lattice results by the BMW Collaboration, the deviation will reduce $\sim 1.6\sigma$ \cite{Borsanyi:2020mff}.
%We should mention results on the hadron vacuum polarization may weaken the necessity of a new physics effect, where they are estimated by recent lattice calculations~\cite{Borsanyi:2020mff,ExtendedTwistedMass:2022jpw,Ce:2022kxy}. 
On the other hand, Refs.~\cite{Crivellin:2020zul,deRafael:2020uif,Keshavarzi:2020bfy} show that the lattice results could conceivably suggest new tensions with the hadron vacuum polarization extracted from $e^+ e^-$ data and the global fits to the electroweak precision observables\footnote{The effect in modifying hadron vacuum polarization for muon $g-2$ and electroweak precision test is also discussed previously in Ref.~\cite{Passera:2008jk}.}.
%%%%%%%%%%%%%%%%%%%%%%%%%%%%%%%%%%%%%%%%%%%%%%%%%%

To get sizable muon $g-2$ at a one-loop level, we would need a chiral flip diagram.
It suggests that we would not obtain one via $\tilde F$.
The new contribution of the Lagrangian is found to be
\begin{align}
-{\cal L}_\nu&= 
 \frac{h}{\sqrt2}{\bar E} \mu  (\eta_{2R}+i\eta_{2I})+ \frac{y_E}{\sqrt2} {\bar\mu} E (\chi_R+i \chi_I)
 + \frac{y'_E}{\sqrt2} {\bar\tau} E (\chi_R+i \chi_I)
 \nn\\
&
 +\frac{A_b v_1}{2\sqrt2}(\eta_{2R}+i\eta_{2I})(\chi_R+i \chi_I)
+{\rm h.c.},
\label{eq:lvs-g2}
\end{align}
where we assume that the mass between the real and imaginary part of $\eta_2$ be the same; $m_{\eta_0}\equiv m_{\eta_{2R}} =m_{\eta_{2I}}$ and $m_{\chi_0}\equiv {\rm Min}[m_{\chi_{2R}}, m_{\chi_{2I}}]$.
Our muon $g-2$ at one-loop level is found as follows~\cite{Lindner:2016bgg, Baek:2016kud}:
\begin{align}
&\Delta a_\mu
=
-{\sqrt2 {m_\mu}} \frac{y_E M_E h A_b v_1}{(4\pi)^2} F_{II}(M_E,m_{\chi_0}, m_{\eta_0}),\\
%%%
%&A_{ij}=\frac{1}{(4\pi)^2} \sum_{a=1}^3 \left(F^\dag_{ja}F_{ai} II(M_a,m_{\eta^-}) { +(|\delta_{ie}G_i^a|^2+|\delta_{i\tau}G_i^a|^2) II(M_a,m_{\chi^-}) }\right),\\
%%%
& F_{II}(M_E,m_{\chi_0}, m_{\eta_0})\simeq 
\int[dx]_4 \frac{x_1+x_2}{[(x_1+x_2) M_E^2 + x_3 m^2_{\chi_0}+ x_4 m^2_{\eta_0}]^2},
\label{eq:damu1}
\end{align}
where $[dx]_4\equiv dx_1dx_2dx_3dx_4\delta(x_1+x_2+x_3+x_4-1)$. 
%%%%%%%%%%%%%%%%%%%%%%%%%%%%%%%%%%%%%%%%%%%%%%%%%%
%\subsection{Anomalous Magnetic Moment of Muon and Electric Dipole Moments}
%%%%%%%%%%%%%%%%%%%%%%%%%%%%%%%%%%%%%%%%%%%%%%%%%%

\subsection{$\tau \to \mu\gamma$ from $y_E$ and $y'_E$}
Due to the terms of $y_E$ and $y'_E$, there is a specific LFV of $\tau \to \mu\gamma$ with no chiral suppressed contribution. The resultant formula is given by
\begin{align}
{\rm BR}(\tau \to \mu\gamma)&\approx \frac{48\pi^3\alpha_{\rm em} C_{32}}{(4\pi)^4 m^2_\tau G^2_F}
(y'_E h A_b v_1 M_E)^2 F_{II}^2(M_E, m_{\eta_0}, m_{\chi_0}).
 \end{align}
Then, $y'_E$ is rewritten in terms of the other terms and experimental constraint as follows:
\begin{align}
y'^2_E
&\lesssim 
\frac{(4\pi)^4 m^2_\tau G^2_F}{48\pi^3\alpha_{\rm em} C_{32}}
\frac{ 4.4 \times 10^{-8}}{(h A_b v_1 M_E)^2 F_{II}^2(M_E, m_{\eta_0}, m_{\chi_0})}.
 \end{align}

\subsection{Dark matter}
We consider several DM candidates; the lightest neutral boson between $\eta_R$ and $\eta_I$, $\chi$ or the lightest of $\psi_{R_i}$ with $i=1,2,3$.
We briefly present the DM candidate of the lightest neutral boson of $\eta_R$ and $\eta_I$, where we concentrate on $\eta_R$ as DM.
In direct detection searches, one might worry that the DM candidate would be ruled out by the spin-independent scattering process mediated by the Z-boson. 
However, it can be evaded if the mass difference between $\eta_R$ and $\eta_I$ {is} bigger than the order 100 keV, since $Z$-boson couples to $\eta_R$ and $\eta_I$, which leads to inelastic scattering and efficiency of the detectors for DM search is much less than $100$keV.

{
If the kinetic term is the only source to explain the correct relic density,
%If the dominant contribution to explain the correct relic density comes from the kinetic term, 
the allowed region is very narrow due to the fixed gauge coupling} and fixed to be around the pole at the half of the neutral Higgs masses, e.g., $\sim 63$ GeV in the case of the SM Higgs, and 534 GeV~\cite{Hambye:2009pw}. On the other hand, when the Yukawa term is dominant to the relic density; $F$ in our case, $F={\cal O}(1)$ is required 
since the cross section is p-wave dominant~\cite{Boehm:2003hm}. 
However, $F$ cannot be so large because the order of our free parameters is 0.01 at most. % as shown in Tables~\ref{bp-tab_nh}, \ref{bp-tab_ih}, \ref{bp-tab_nh_i}. 
%
% the additional mixing from unitary mixing $V$ that diagonals the neutral heavier fermion ${\cal M}_N$.
%%%
Fermionic DM candidate $\psi_{R_1}$ has annihilation channels only through $F$ that is also $p$-wave dominant.
Thus, we cannot explain the correct relic density of the fermionic DM due to too small a thermally averaged cross-section.
%We now move forward to another DM candidate $\chi$.

We now move forward to another DM candidate $\chi$ where we assume a complex scalar DM candidate supposing the mass difference between $\chi_R$ and $\chi_I$ is small enough to consider the single complex DM.
Here, we simply suppose that the relic density would be explained by a combination of Yukawa couplings $y_E$ and $g_E$ that give $s$-wave dominant as can be seen in muon $g-2$, by assuming all the other interactions from the Higgs potential is negligibly small. 
The relevant Lagrangian originates from the same one of muon $g-2$.
Then, the {cross-section} in terms of relative velocity $v_{\rm rel}$ is given by~\cite{Chiang:2017zkh}
\begin{align}
\sigma v_{\rm rel}&\approx 
 \frac{1}{4\pi} 
\frac{|A_b v_1 y_E M_E h|^2}{(m_X^2 + M_E^2)^2(m_X^2 - m_{\eta_0}^2)^2} 
+\frac{(Ab v _1 h M_E y'_E)^2}{4(M_E^2+m_X)^2(m_{\eta_0}^2+m^2_X)^2\pi}\nn\\
%%%
&+\frac{|y_E|^4}{96\pi} 
\frac{m_X^2}{(m_X^2 + M_E^2)^2} v^2_{\rm rel}
+\frac{|y'_E|^2 (|y_E|^2+|y'_E|^2)}{96\pi} 
\frac{m_X^2}{(m_X^2 + M_E^2)^2} v^2_{\rm rel}
 + {\cal O} (v^4_{\rm rel}),
\end{align}
where $\chi$ is considered as complex, assuming that $m_\mu\ll m_\chi,  M_E$, and the approximation method is used to expand the relative velocity of the DM $v_{\rm rel}^2\sim 0.3$.
The cross-section should lie within the range of [1.78-1.97]$\times 10^{-9}$
GeV$^{-2}$ at 2$\sigma$ C.L. that leads to the correct relic density $\Omega_{\rm DM} h^2 \sim0.12$.
However, we will use a more relaxed bound in our numerical analysis; $[1.5-3.5]\times 10^{-9}$ GeV$^{-2}$.
In DM direct detection, we expect that it would not give us so stringent bounds, since the $\chi$ does not directly couple to the quarks in SM at the tree level. The same analysis has been studied in Ref.~\cite{Schmidt:2012yg}, however, it was done at a one-loop level.

%%%%%%%%%%%%%%%%%%%%%%%%%%%%%%%%%%%%%%%%%%%%%%%%%%
\section{Numerical analysis in lepton sector \label{sec:NAl}}
%%%%%%%%%%%%%%%%%%%%%%%%%%%%%%%%%%%%%%%%%%%%%%%%%%
In this Section, we present our numerical analysis focusing on the neutrino oscillation data.
Since the DM and muon $g-2$ have independent parameters such as $h, y_E$, and $M_E$, we will search for these observed parameters separately after the analysis of the lepton sector.
We use the following ranges of input parameters below,
\begin{align}
&\{b_\eta/a_\eta, c_\eta/a_\eta, |\epsilon|, |r_1|, |r_2| \} \in [10^{-3},10^3], \nonumber\\
%\ y_E \in [0.1,\sqrt{4\pi}],\nn\\
& m_{\chi_R} \in [10^2,10^5]{\rm GeV}, \  m_{\chi_I} \in [m_{\chi_R},m_{\chi_R}+50]{\rm GeV}, 
 \ [m_{\eta_1} (\simeq m_{\eta^\pm}) ,M_0]
 \in [m_{\chi_R} ,10^7] \ {\rm GeV},
\end{align}
where we work on the fundamental region of $\tau$, the mass degeneracy among $\eta$ implicitly contains to evade constraints of the oblique parameters in our numerical analysis, and $\epsilon, r_1,r_2$ are complex that appear in the case of $k_N=3$ only.
Under these regions, we randomly scan those input parameters and search for the allowed regions by imposing the constraints of the neutrino oscillation data and LFVs. Next, we will show some plots in terms of the classification of $\Delta\chi^2$ analysis within the range of 2$\sigma$ which is represented by green color, 2$\sigma$-3$\sigma$ which is represented by a yellow color, and 3$\sigma$-5$\sigma$ which is represented by red color, referring to NuFit 5.1~\cite{Gonzalez-Garcia:2021dve}, where we have taken the square root for $\Delta\chi^2$.
In the present work, we adopt the accuracy of $\Delta\chi^2$ for five well known observables such as $\Delta m_{\rm atm}^2$, $\Delta m_{\rm sol}^2$, $\sin^2 \theta_{12}$, $\sin^2 \theta_{23}$, and $\sin^2 \theta_{13}$ {\it satisfying the bounds on LFVs in Eqs~(\ref{eq:exp-lfvs})}. 
 
{%\color{red}
%{\Large Only the figures are updated!!!}

\subsection{Case of $k_N=1$\label{subsec:kn=1}}
At first, we show chi-square analysis in the case of $k_N=1$.

\subsubsection{$NH$ \label{sec:NA1}}

%------------------------------------------------------------------
\begin{figure}[htbp]
  \includegraphics[width=77mm]{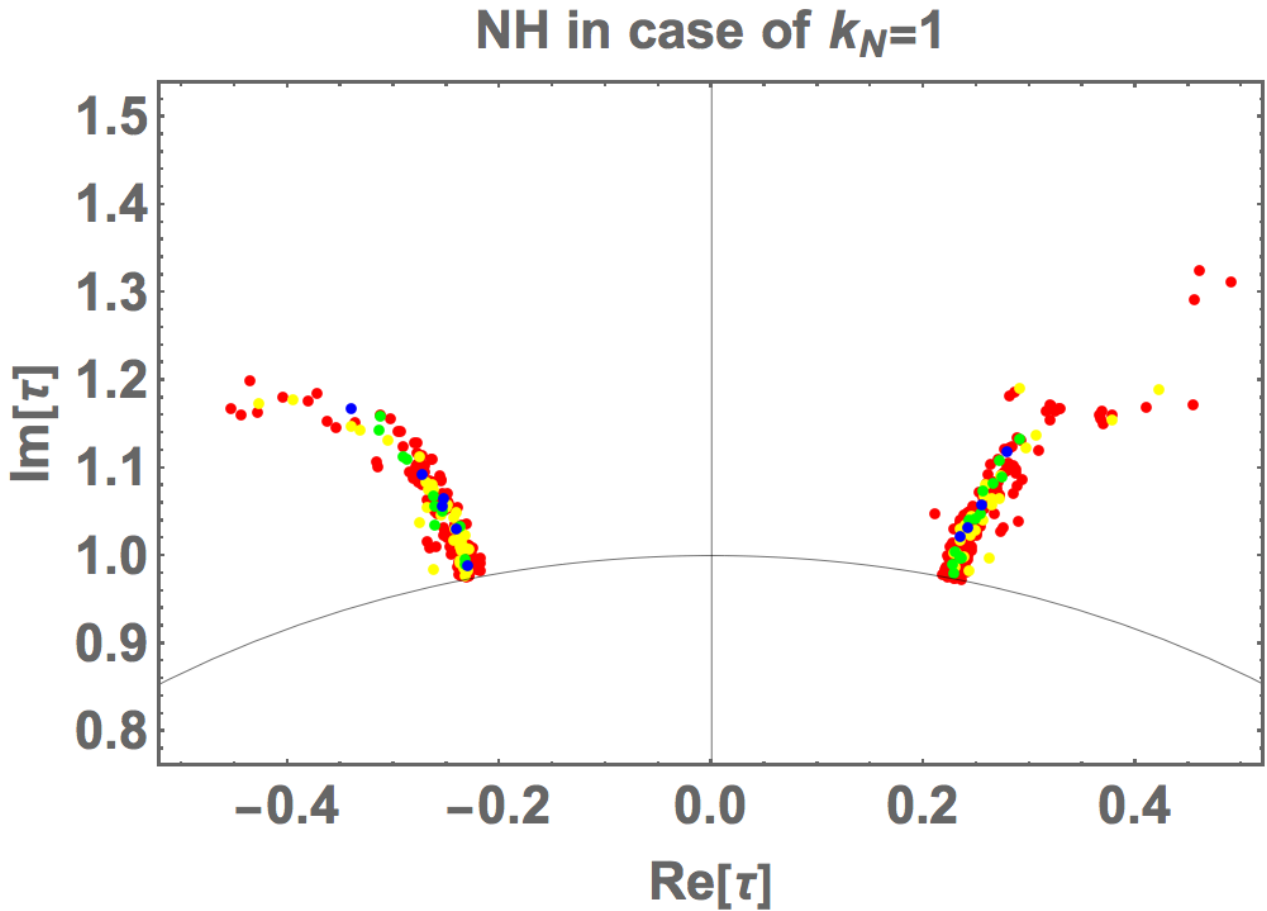}
  \caption{The scatter plots for the real $\tau$ and imaginary $\tau$ in case of NH and $k_N=1$.
    In the $\Delta\chi^2$ analysis, the green color represents the range of $0-2$, the yellow color is for $2-3$, and the red color is for $3-5$ of $\sqrt{\Delta\chi^2}$. The black solid line is the boundary of the fundamental domain at $|\tau|=1$.}
  \label{fig:tau_nh-1}
\end{figure}
%%%
In Fig.~\ref{fig:tau_nh-1}, we show the scatter plots of the real $\tau$ and imaginary $\tau$ in the case of NH and $k_N=1$.
The solid line is the fundamental domain boundary at $|\tau|=1$.

\begin{figure}[htbp]
  \includegraphics[width=53.5mm]{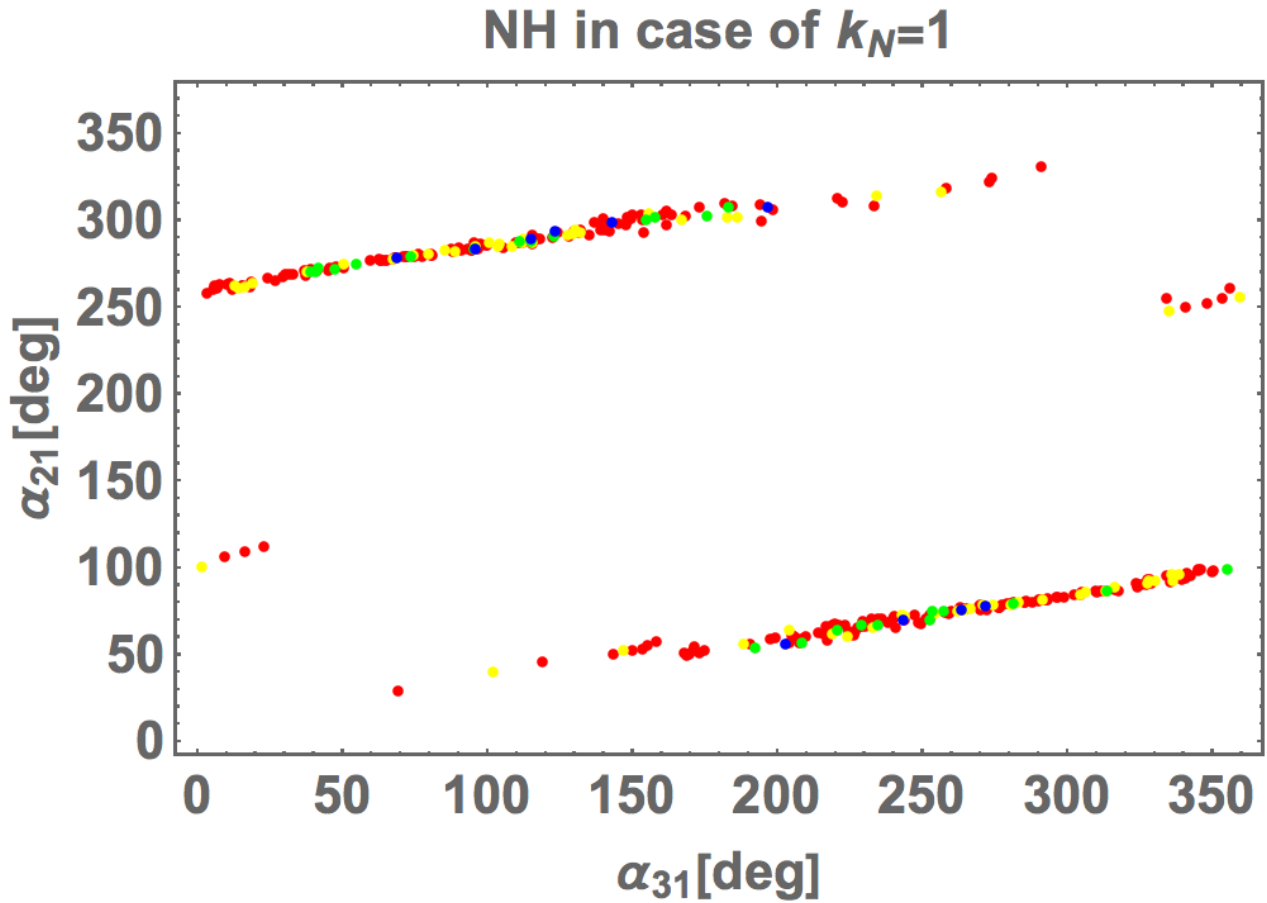}
  \includegraphics[width=53.5mm]{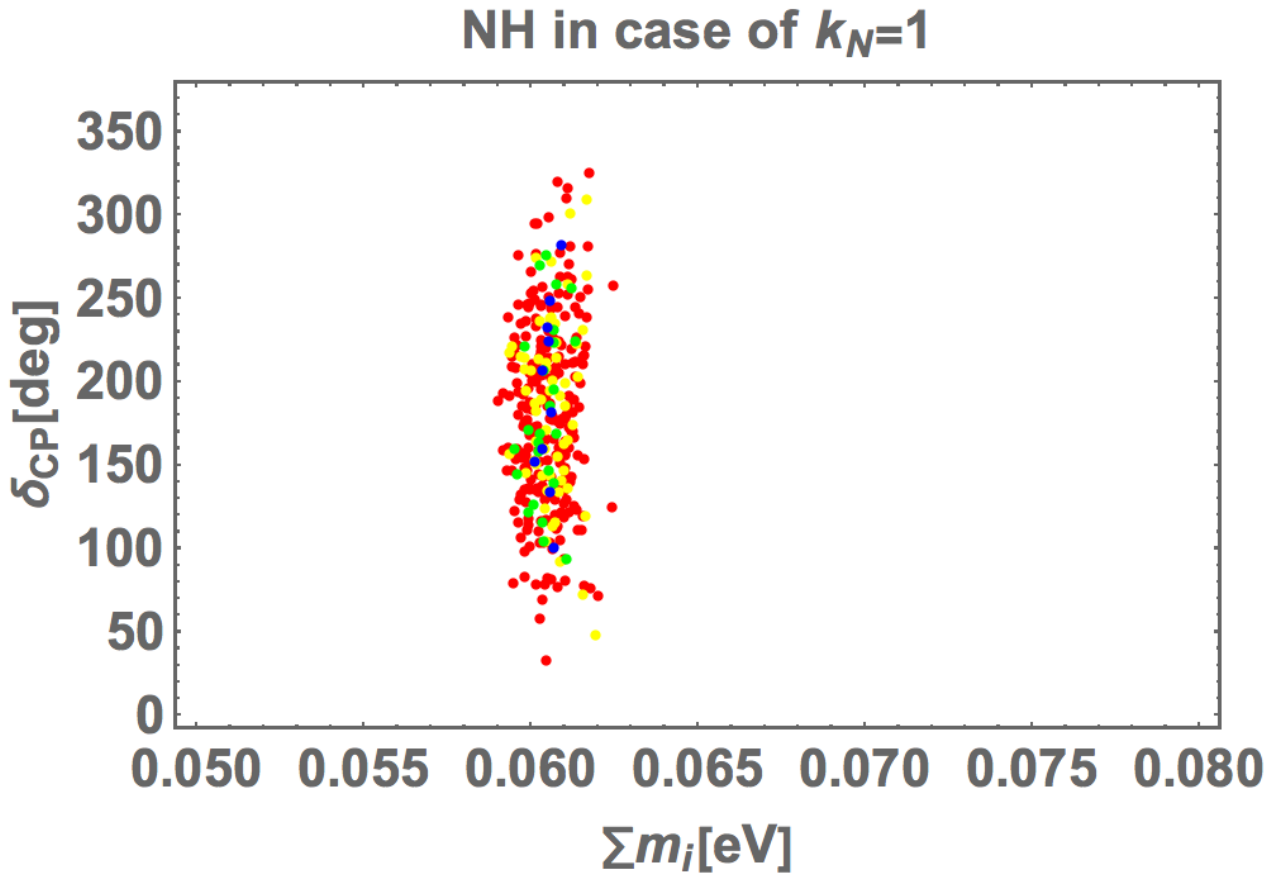}
  \includegraphics[width=53.5mm]{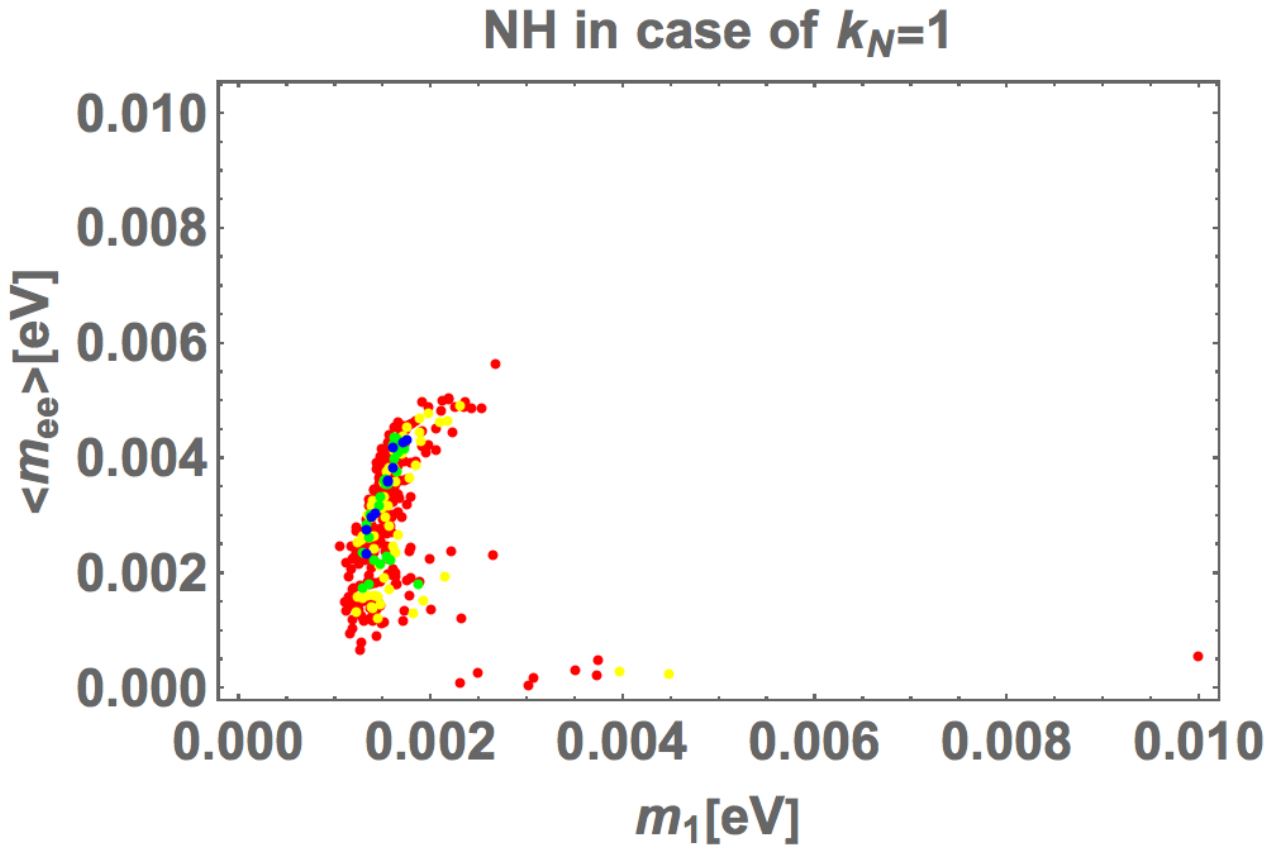}
  \caption{The scatter plots for the Majorana phases $\alpha_{31}$ and $\alpha_{21}$ (left panel), the summation of active neutrino mass eigenstates and Dirac CP phase (middle panel), and the lightest mass of active neutrinos ($m_1$) and neutrinoless double beta decay ($\langle m_{ee}\rangle$) (right panel) in case of NH and $k_N=1$. The color representations are the same as in Fig.~\ref{fig:tau_nh-1}.}
  \label{figs_obs-nh-1}
\end{figure}

In Fig.~\ref{figs_obs-nh-1}, we show the scatter plots for the Majorana phases $\alpha_{31}$ and  $\alpha_{21}$ (left panel), the summation of active neutrino mass eigenstates and Dirac CP phase (middle panel), and the lightest mass of active neutrinos ($m_1$) and neutrinoless double beta decay ($\langle m_{ee}\rangle$) (right panel) in case of NH. The color representations are the same as in Fig.~\ref{fig:tau_nh-1}.
%%%
From the left panel, we find linear correlations between Majorana phases; any value is allowed for $\alpha_{31}$ while $\alpha_{21}=[0-120, 250-330]$ deg.
%%%
From the middle panel, allowed regions of $\sum_{i}m_i$ are localized at nearby $60$ meV, while any value of $\delta_{CP}$ is allowed except the region at nearby zero. Whole the region is within the bound on cosmological constant $\sum m_i\le120$ meV.
%%%
From the right panel, {the main} allowed regions are localized at $m_1=[1-3]$ meV with $\langle m_{ee}\rangle=[0.5-5.5]$ meV even though there are a few sporadic points up to 10 meV of $m_1$ are allowed.

%%%%%%%%%
\begin{figure}[htbp]
  \includegraphics[width=77mm]{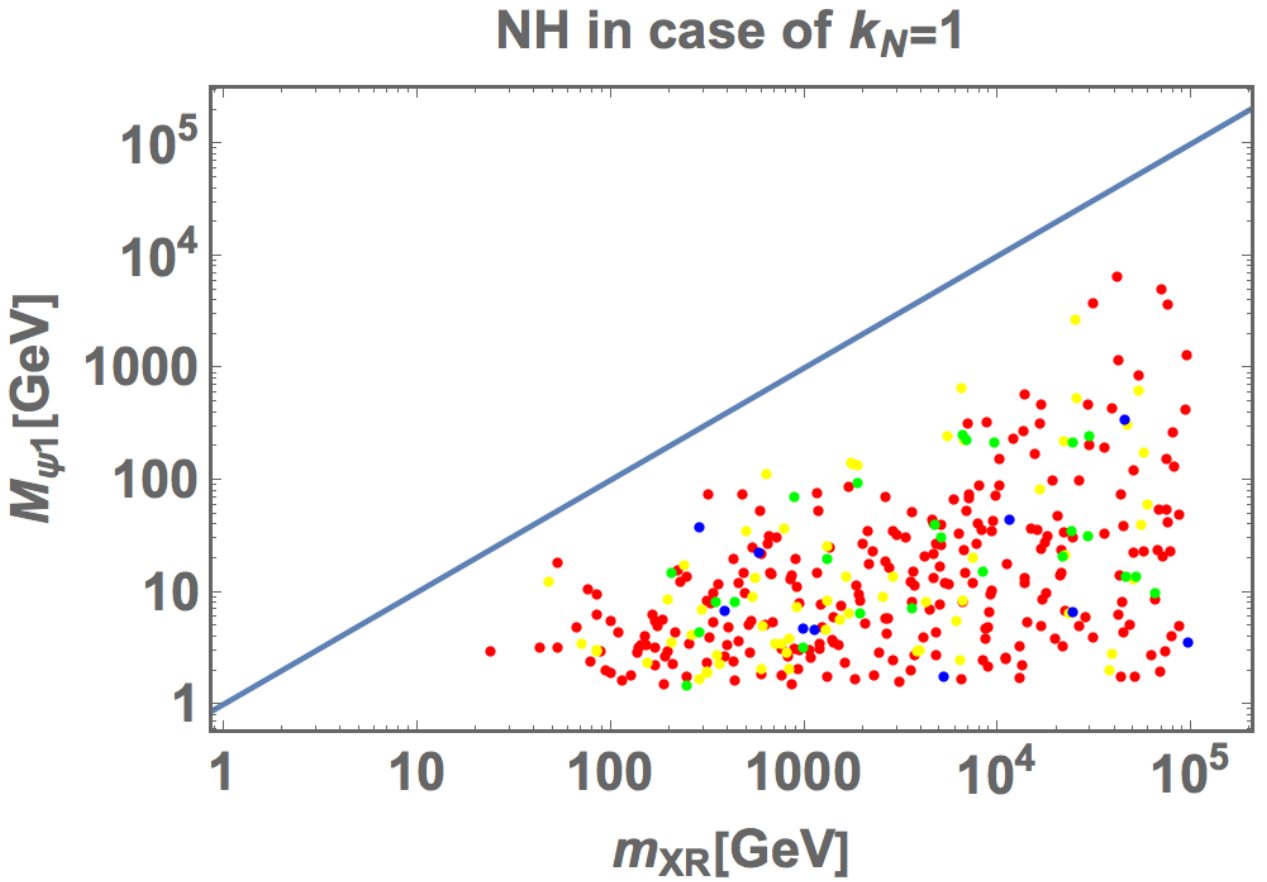}
   \caption{The scatter plots for $m_{\chi_R}$ and $M_{\psi_1}$, where the color representations are the same as in Fig.~\ref{fig:tau_nh-1} and the blue line shows $m_{\chi_R}=M_{\psi_1}$. }
  \label{figs_mdm-nh-1}
\end{figure}
In Fig.~\ref{figs_mdm-nh-1}, we show the scatter plots for the DM candidates $M_1$ and $m_{\chi_R}$ in the case of NH. 
Here, the color representations are the same as in Fig.~\ref{fig:tau_nh-1}.
This figure shows that all the mass of $\chi_{R}$ is heavier than the one of $\psi_1$, therefore $\chi_{R}$ cannot be a DM candidate.
%%%

\subsubsection{$IH$ \label{sec:NA1}}
%

%------------------------------------------------------------------
\begin{figure}[htbp]
  \includegraphics[width=77mm]{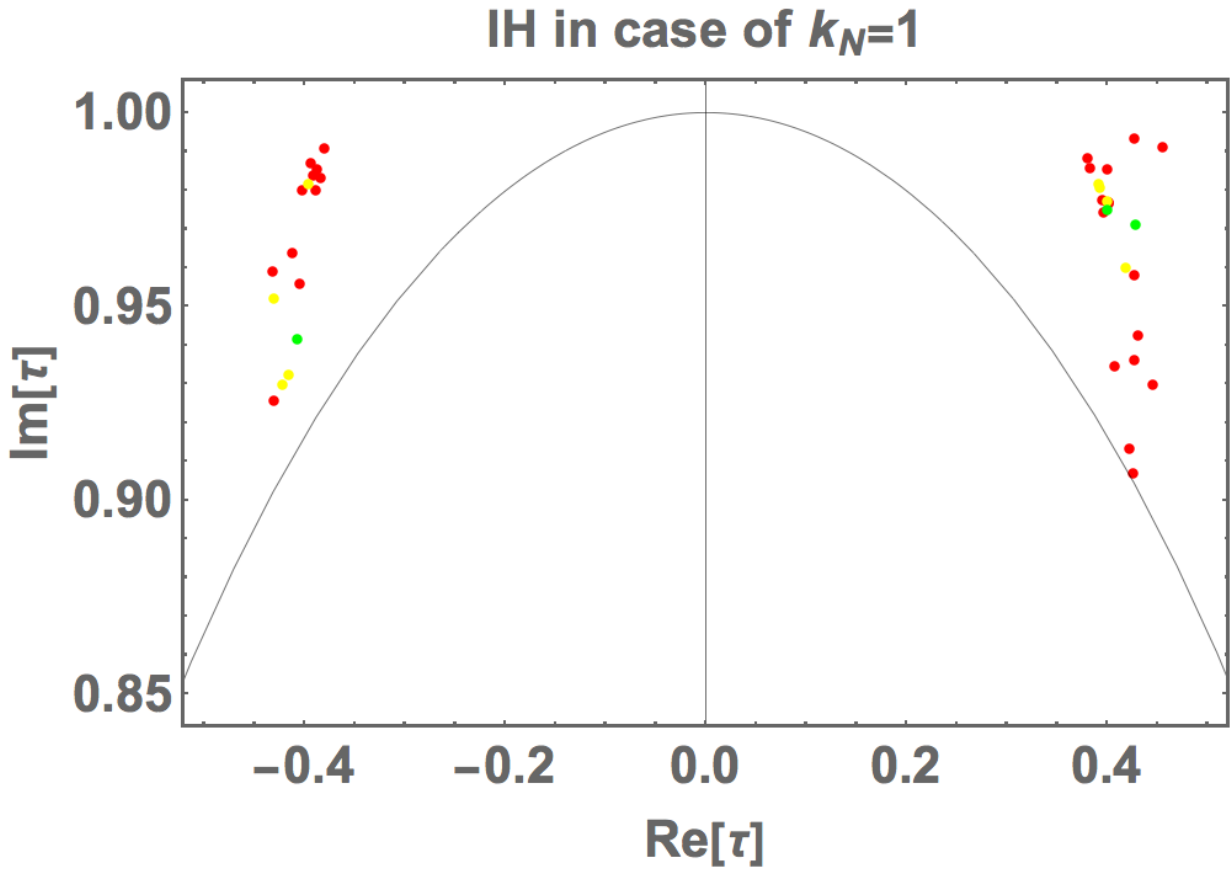}
  \caption{The scatter plots for the real $\tau$ and imaginary $\tau$ in case of IH and $k_N=1$.
Here, the color representations and black line {is} the same as in Fig.~\ref{fig:tau_nh-1}.}
  \label{fig:tau_ih-1}
\end{figure}
%%%
In Fig.~\ref{fig:tau_ih-1}, we show the scatter plots of the real $\tau$ and imaginary $\tau$ in {the} case of IH and $k_N=1$.
Here, the color representations and black line {is} the same as in Fig.~\ref{fig:tau_nh-1}.
Here, there are no solutions within the range of $2\sigma$. This figure shows that the allowed region would be near a fixed point $\tau=\pm\frac12+i\frac{\sqrt3}{2}$.

\begin{figure}[htbp]
  \includegraphics[width=53.5mm]{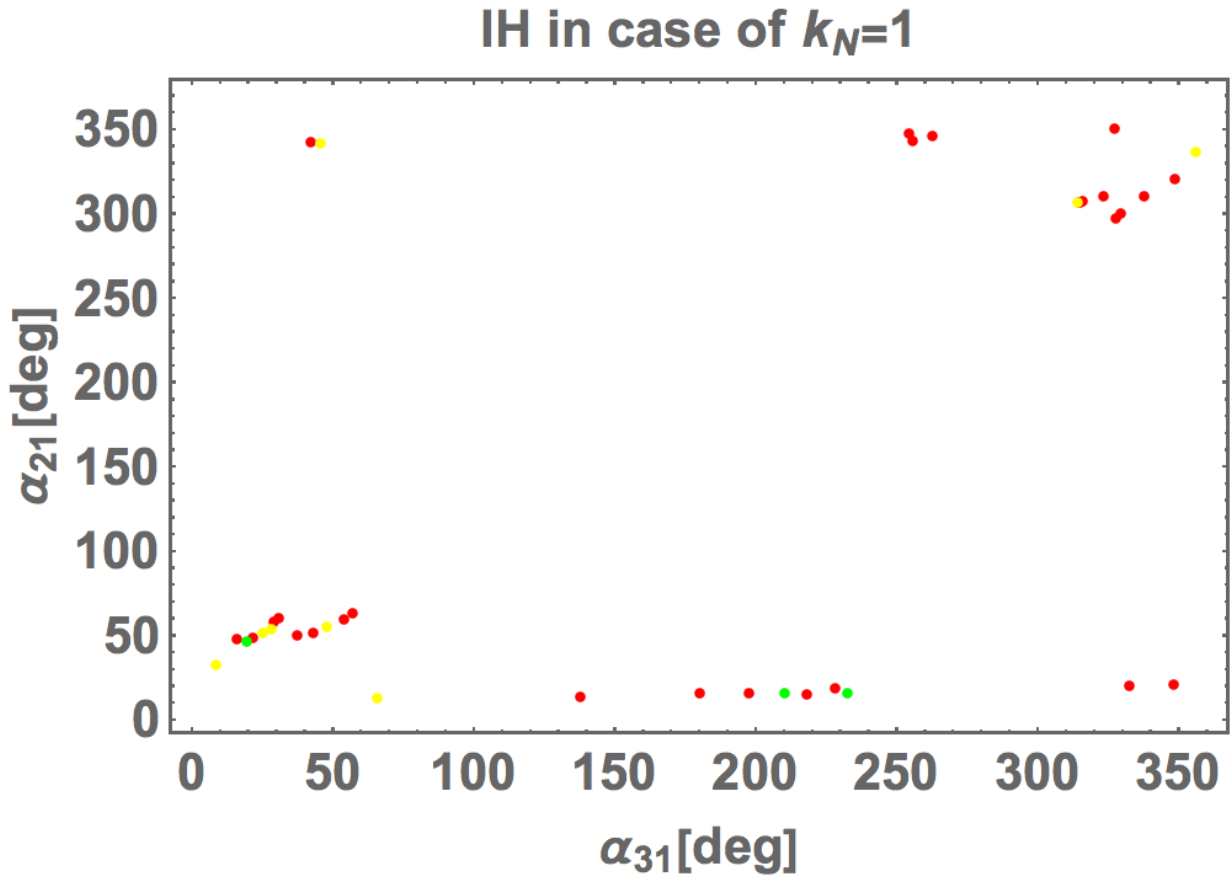}
  \includegraphics[width=53.5mm]{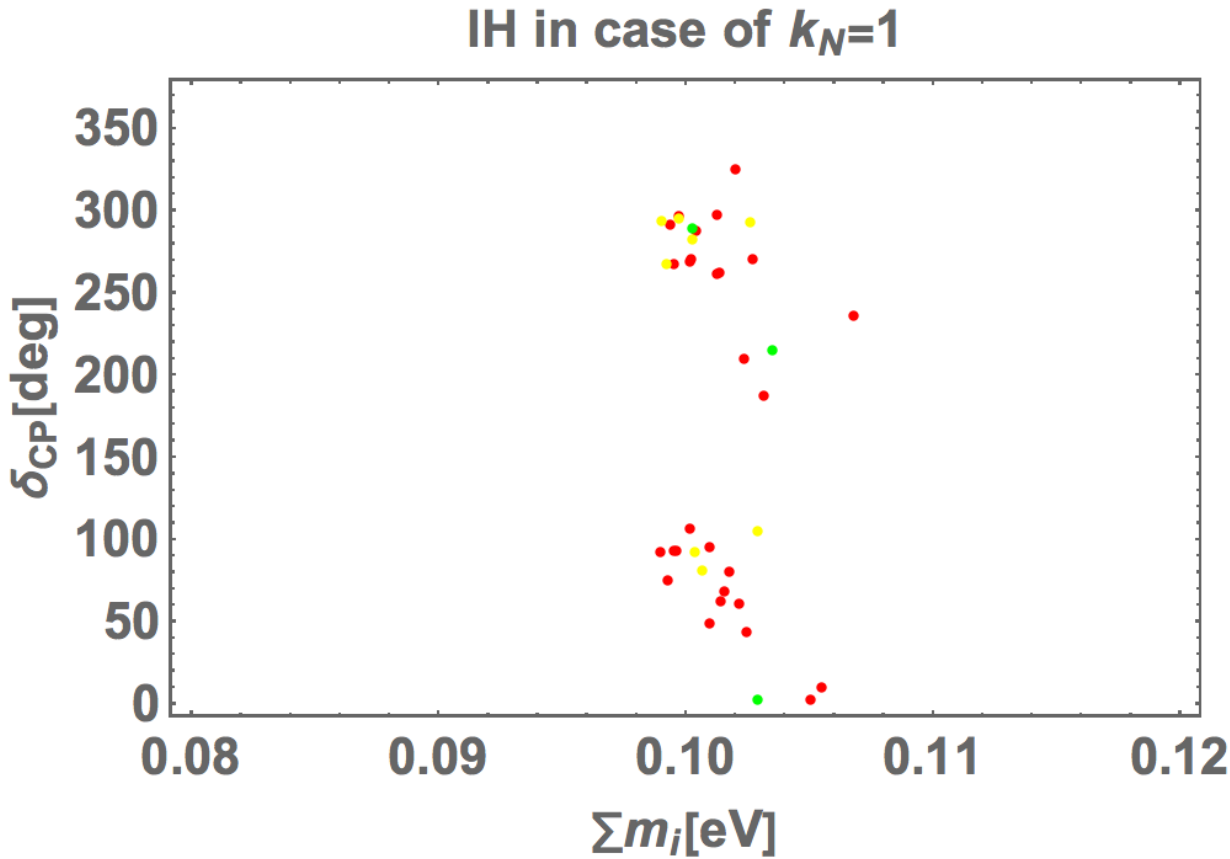}
  \includegraphics[width=53.5mm]{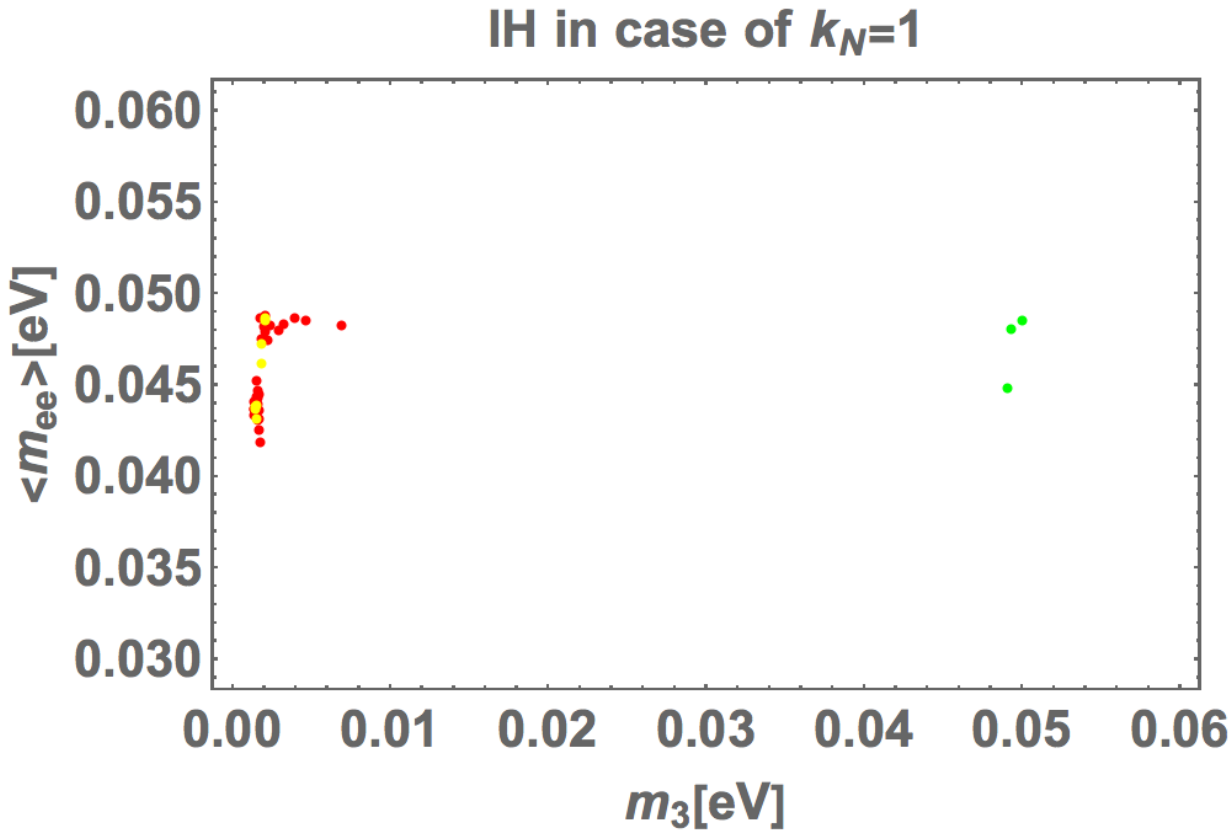}
  \caption{The scatter plots for the Majorana phases $\alpha_{31}$ and  $\alpha_{21}$ (left panel), the summation of active neutrino mass eigenstates and Dirac CP phase (middle panel), and the lightest mass of active neutrinos ($m_3$) and neutrinoless double beta decay ($\langle m_{ee}\rangle$) (right panel) in case of IH. The color representations are the same as in Fig.~\ref{fig:tau_nh-1}.}
  \label{figs_obs-ih-1}
\end{figure}
%%%%%%%%%
In Fig.~\ref{figs_obs-ih-1}, we show the scattering plots in the case of IH that is {the same} as the case of NH. The color representations are the same as in Fig.~\ref{fig:tau_nh-1}.
From the left panel, we find that almost any value is allowed for $\alpha_{31}$, while $\alpha_{21}$ is localized at nearby zero.
%%%
From the middle panel, we find $\sum_{i}m_i$ is localized at nearby 0.1 eV while $\delta_{CP}=[0-100, 200-330]$ deg. 
All allowed regions are within the bound of cosmological constant $\sum m_i\le120$ meV.
%%%
From the right panel, we find that there are two allowed regions for $m_3$; 5 meV or 50 eV, while $\langle m_{ee}\rangle=[42-49]$ meV.

%%%%%%%%%
\begin{figure}[htbp]
  \includegraphics[width=77mm]{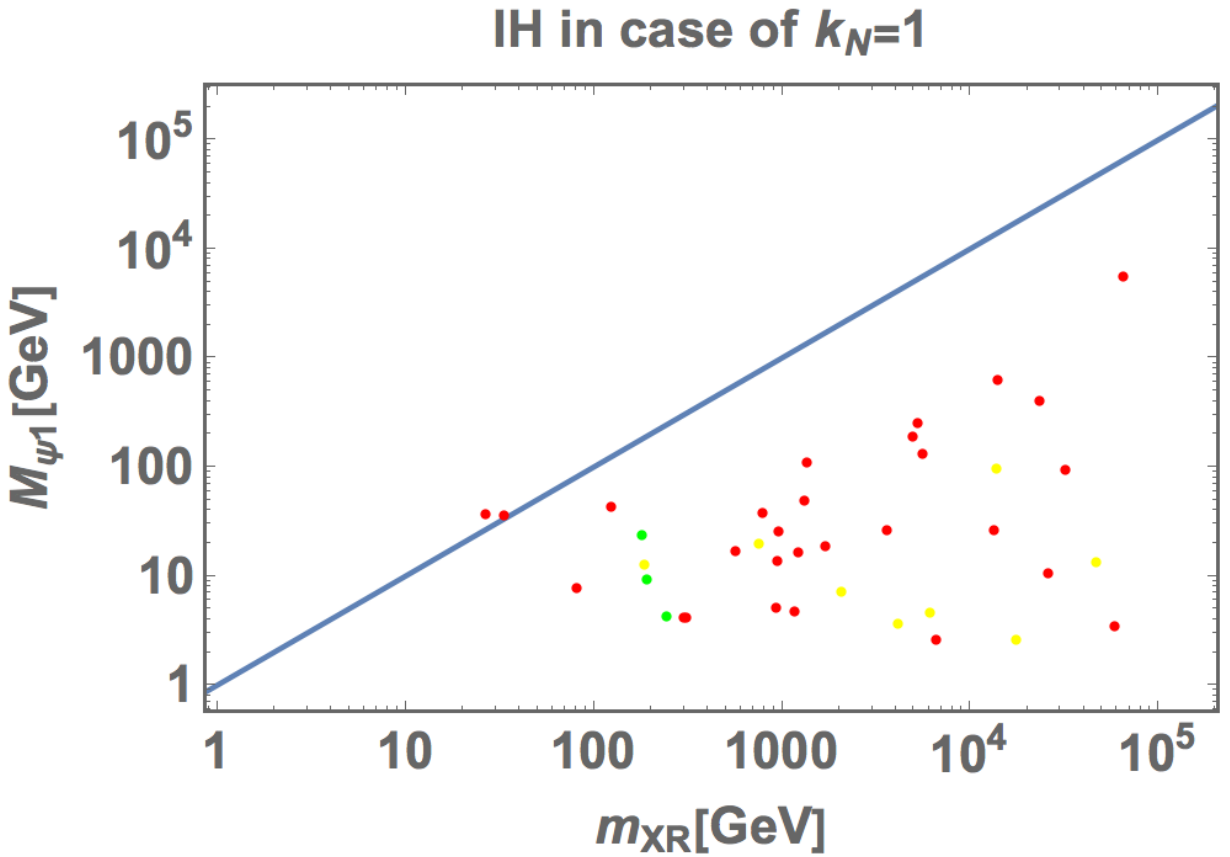}
   \caption{The scatter plots for $m_{\chi_R}$ and $M_{1}$. The color representations are the same as in Fig.~\ref{fig:tau_nh-1} and the blue line shows $m_{\chi_R}=M_{\psi_1}$.}
  \label{figs-mdm_ih-1}
\end{figure}
In Fig.~\ref{figs-mdm_ih-1}, we show the scatter plots for the DM candidates $M_1$ and $m_{\chi_R}$ in {the} case of IH. 
Here, the color representations are the same as in Fig.~\ref{fig:tau_nh-1}.
This figure shows that the mass of $\chi_{R}$ is heavier than the one of $\psi_1$ except for two points, therefore the situation is almost similar to the case of NH.~\footnote{Although we have two points satisfying $m_{\chi_R}\le M_1$, each of field is degenerate. It implies that we have to consider two components of the DM scenario or coannihilation processes whose analysis would be rather complicated. Thus, we do not consider such a situation, too.}

%%%
\subsubsection{Summary of $k_N=1$}
Even though we have several predictions on neutrino oscillation data for both the cases of NH and IH,
we have no DM candidate on $\chi_R$ as discussed in the DM subsection.
Thus, we extend our model by increasing the modular weight in the next subsection.

\subsection{Case of $k_N=3$\label{subsec:kn=3}}
Second, we show chi-square analysis in the case of $k_N=3$.
Due to the larger value of the modular weight, we have three more free parameters that would contribute to our DM candidate $\chi_R$ {working} well in the neutrino analysis. Thus, we further impose the constraint $m_{\chi_R}\le M_1$ below.

\subsubsection{$NH$ \label{sec:NA1}}

%------------------------------------------------------------------
\begin{figure}[htbp]
  \includegraphics[width=77mm]{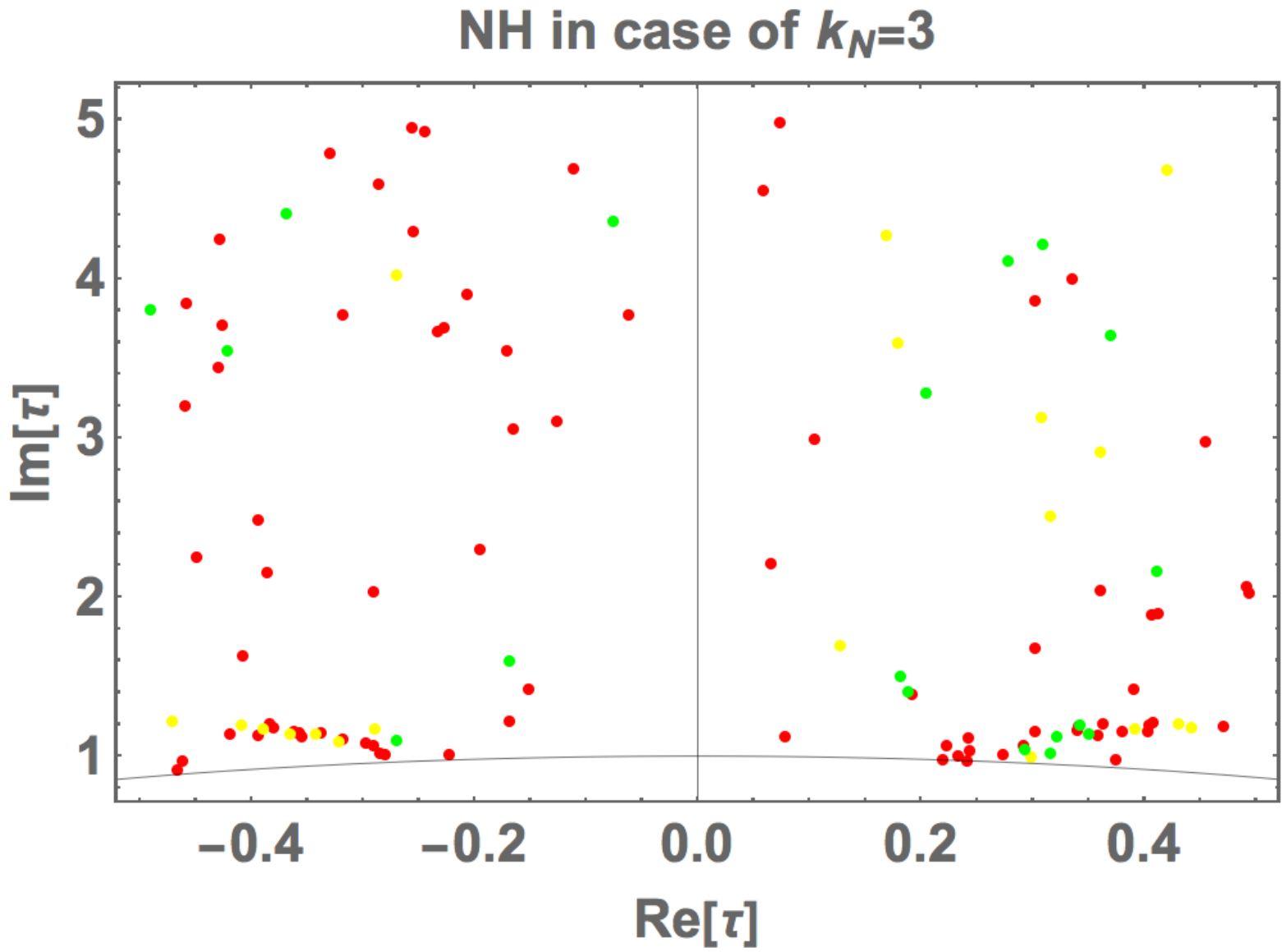}
  \caption{The scatter plots for the real $\tau$ and imaginary $\tau$ at nearby the fixed point $\tau=\omega$ in NH.
    In the $\Delta\chi^2$ analysis, the green color represents the range of $0-2$, the yellow color is for $2-3$, and the red color is for $3-5$ of $\sqrt{\Delta\chi^2}$. The black solid line is the boundary of the fundamental domain at $|\tau|=1$.}
  \label{fig:tau_nh-2}
\end{figure}
%%%
In Fig.~\ref{fig:tau_nh-2}, we show the scatter plots of the real $\tau$ and imaginary $\tau$ in {the} case of NH. 
The solid line is the fundamental domain boundary at $|\tau|=1$.
The allowed region lies on whole the place.

\begin{figure}[htbp]
  \includegraphics[width=53.5mm]{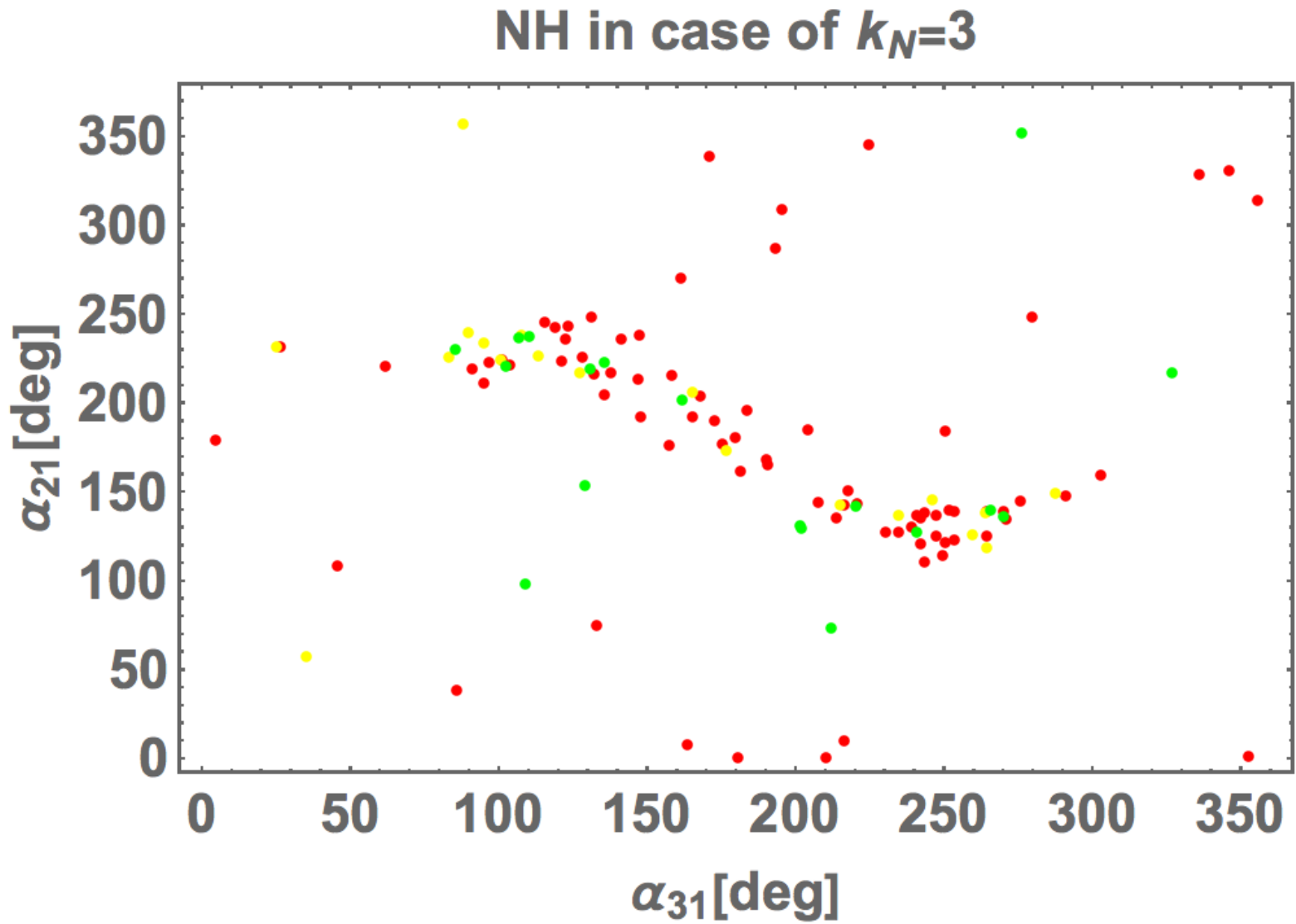}
  \includegraphics[width=53.5mm]{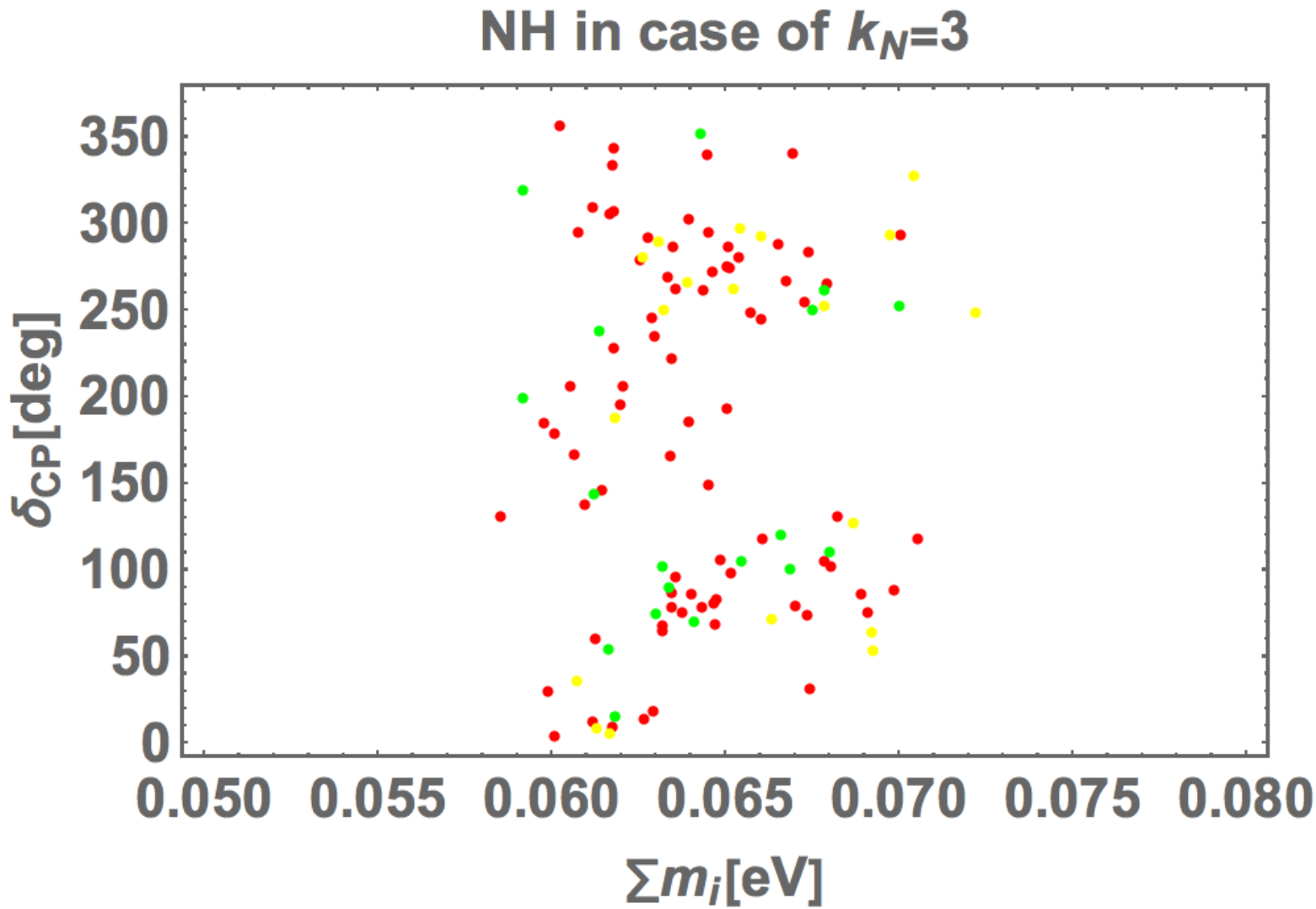}
  \includegraphics[width=53.5mm]{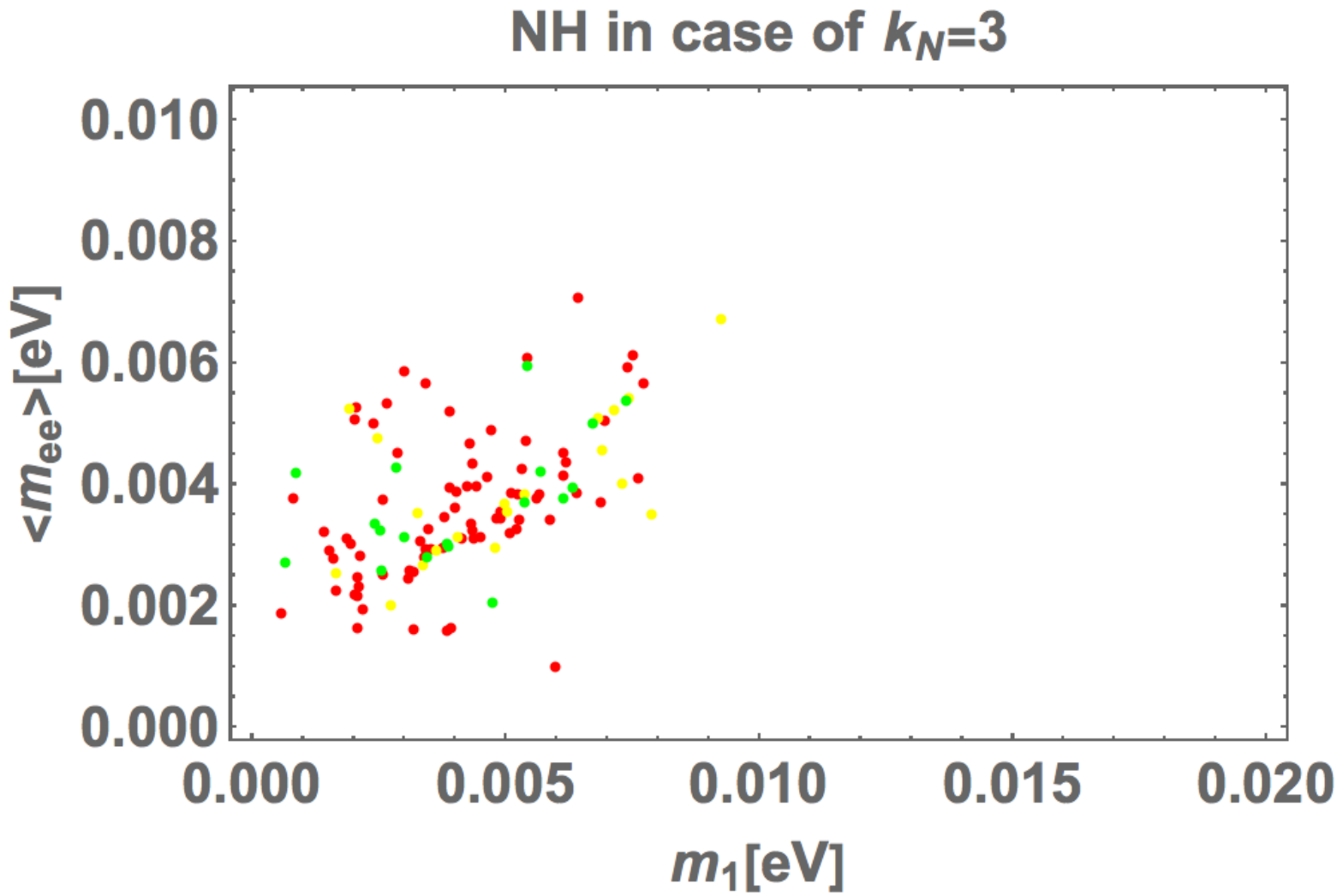}
  \caption{The scatter plots for the Majorana phases $\alpha_{31}$ and  $\alpha_{21}$ (left panel), the summation of active neutrino mass eigenstates and Dirac CP phase (middle panel), and the lightest mass of active neutrinos ($m_1$) and neutrinoless double beta decay ($\langle m_{ee}\rangle$) (right panel) in case of NH. The color representations are the same as in Fig.~\ref{fig:tau_nh-1}.}
  \label{figs-obs_nh-2}
\end{figure}
In Fig.~\ref{figs-obs_nh-2}, we show the scatter plots for the Majorana phases $\alpha_{31}$ and  $\alpha_{21}$ (left panel), the summation of active neutrino mass eigenstates and Dirac CP phase (middle panel), and the lightest mass of active neutrinos ($m_1$) and neutrinoless double beta decay ($\langle m_{ee}\rangle$) (right panel) in case of NH. The color representations are the same as in Fig.~\ref{fig:tau_nh-1}.
%%%
From the left panel, we find correlations between Majorana phases. Even though any values are allowed, there is {a tendency}
 $\alpha_{31}=[70-300]$ deg and  $\alpha_{31}=[100-250]$ deg. 
%From the left panel, we find that there are four localized regions; $\alpha_{31}=[80-125, 230-285]$ deg and $\alpha_{21}=[0-20, 340-360]$ deg.
%%%
From the middle panel, allowed regions of $\sum_{i}m_i$ is $[60-72]$ meV, while all the region is allowed for $\delta_{CP}$. 
Whole the region is within the bound on cosmological constant $\sum m_i\le120$ meV.
%%%
From the right panel, allowed regions are $m_1=[0-9]$ meV and $\langle m_{ee}\rangle=[0-7]$ meV.
%%%

%%%%%%
\begin{figure}[htbp]
  \includegraphics[width=77mm]{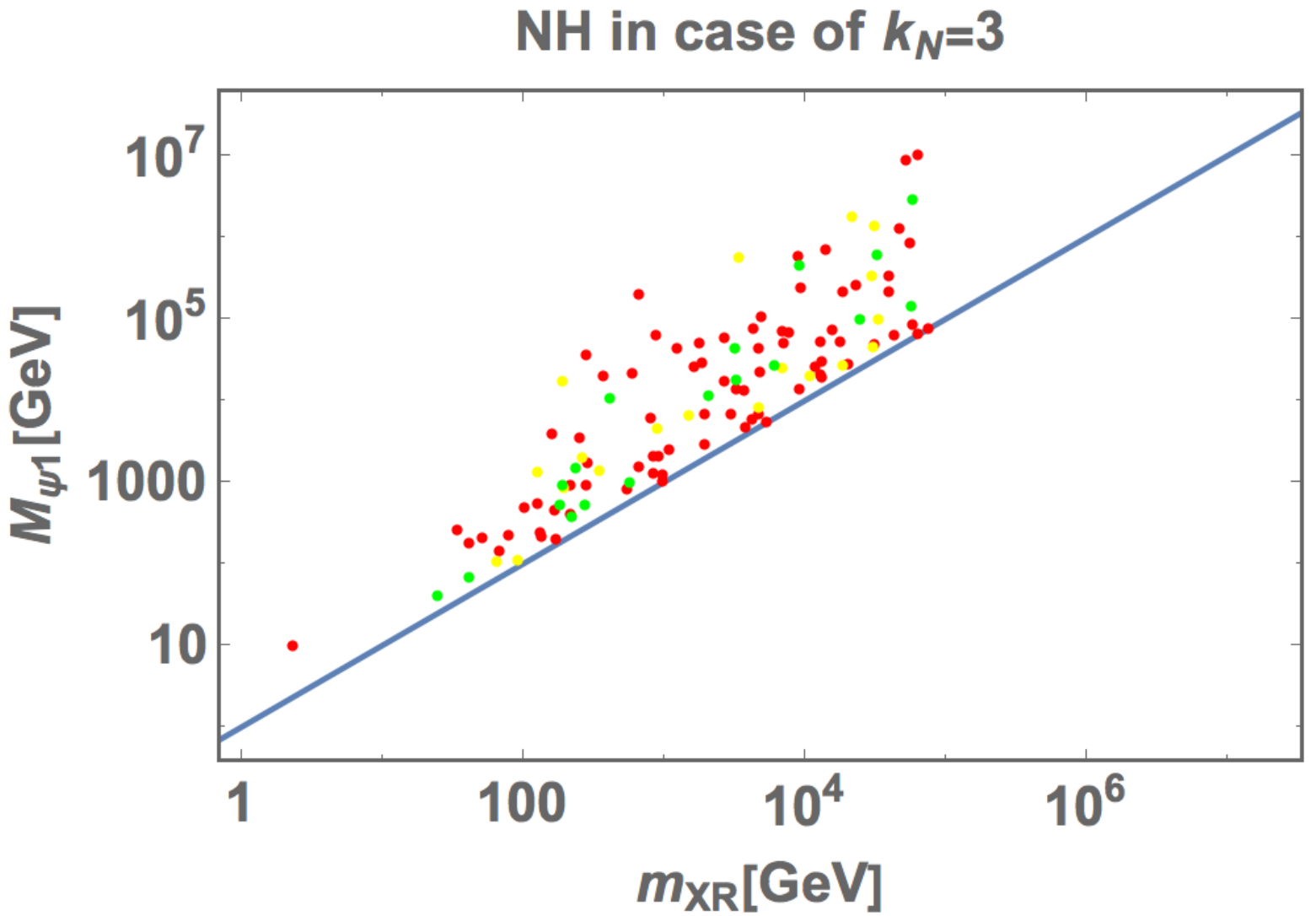}
   \caption{The scatter plots for $m_{\chi_R}$ and $M_{\psi_1}$. The color representations are the same as in Fig.~\ref{fig:tau_nh-1} and the blue line shows $m_{\chi_R}=M_{\psi_1}$.}
  \label{figs-mdm_nh-2}
\end{figure}
In Fig.~\ref{figs-mdm_nh-2}, we show the scatter plots for the DM candidates $M_1$ and $m_{\chi_R}$ in {the case} of NH. 
Here, the color representations are the same as in Fig.~\ref{fig:tau_nh-1}.
This figure represents that DM $\chi_{R}$ works well. 
%%%

%%%%%%%%%%%%%%%%%%%%%%%%%%%%%%%%%%%%%%%%%%%%%%%%%%%
\begin{table}[h]
	\centering
	\begin{tabular}{|c|c|} \hline 
			\rule[14pt]{0pt}{0pt}
%Lepton		&  NH($\tau\approx 1.75 i$) & IH($\tau\approx i$)  & IH($\tau\approx 1.76 i$)\\  \hline
 		&  NH  \\  \hline
			\rule[14pt]{0pt}{0pt}
		$\tau$ & $-0.456 + 4.23 i$      \\ \hline
		\rule[14pt]{0pt}{0pt}
		%%%
%		
		$[b_\eta/a_\eta,c_\eta/a_\eta]$ & $[-2.87629,\ -29.7075]$      \\ \hline
		\rule[14pt]{0pt}{0pt}
		$[ r_1, r_2, \epsilon]$ & $[-9.55 + 53.3 i,\ 683 - 20.6i ,\ 53.3 - 0.261i]$      \\ \hline
		\rule[14pt]{0pt}{0pt}
%		$c_\eta$  &  $0.00767509 - 0.0313634 i$ & $0.0013744 - 0.0000470787 I$ & $-0.00227049 + 0.00350111 I$\\
%		$\epsilon$ & $-0.0619 + 0.00436  i$ & $-0.0626 - 0.00268 i$     \\ \hline
%		\rule[14pt]{0pt}{0pt}
		%%%
%		$a$  &  $0.000156626 + 0.000114523 i$ & $3.04135 + 0.98336 I$ & $0.00853417 + 0.0246743 I$\\
		$M_0/{\rm GeV}$ & $111.722$    \\ \hline
		\rule[14pt]{0pt}{0pt}
%		$b$  &  $-0.0173463 + 0.0012792 i$ & $0.000143616 + 0.000153805 I$ & $-0.0960243 + 0.0408818 I$\\
		$[M_1,M_2,M_3]/ {\rm GeV}$  &  $[2.54\times10^4, 2.54\times10^4, 2.55\times10^4] $     \\ \hline
		\rule[14pt]{0pt}{0pt}
		%$b$  &  $-0.0173463 + 0.0012792 i$ & $0.000143616 + 0.000153805 I$ & $-0.0960243 + 0.0408818 I$\\
		$[m_{\chi_R},m_{\chi_I},m_{\eta_1}]/ {\rm GeV}$  &  $[182,\ 183,\ 187] $  %$[182.,\ 232,\ 187] $  
		  \\ \hline
		\rule[14pt]{0pt}{0pt}
%		$c$  &  $-0.000300709 - 0.000100567 i$ & $0.005563 - 0.0021227 I$ & $-1.78167 + 4.54256 I$\\
		$\kappa/ {\rm GeV}$  &  $8.22\times10^{-5} $   \\ \hline
		\rule[14pt]{0pt}{0pt}
%		$b'$  &  $0.000137596 - 0.0000163875 i$ & $0.0000236483 + 0.0000158969 I$ & $-0.000105682 + 0.000106792 
		$\Delta m^2_{\rm atm}$  &  $2.52\times10^{-3} {\rm eV}^2$   \\ \hline
		\rule[14pt]{0pt}{0pt}
%		$c'$  &  $-0.213667 - 0.271705 i$ & $-0.153721 + 0.0313641 I$ & $-3.40695 \times 10^{-6} - 0.0000598624 I$\\
		$\Delta m^2_{\rm sol}$  &  $7.18\times10^{-5} {\rm eV}^2$        \\ \hline
		\rule[14pt]{0pt}{0pt}
%		$\sin^2\theta_{12}$ & $ 0.322231$& $0.28036$ & $0.33157$\\
		$\sin\theta_{12}$ & $ 0.569$   \\ \hline
		\rule[14pt]{0pt}{0pt}
%		$\sin^2\theta_{23}$ &  $ 0.563489$& $0.463487$ & $0.579596$\\
		$\sin\theta_{23}$ &  $ 0.765$     \\ \hline
		\rule[14pt]{0pt}{0pt}
%		$\sin^2\theta_{13}$ &  $ 0.0235092$&$0.0240532$ & $0.0218055$\\
		$\sin\theta_{13}$ &  $ 0.149$    \\ \hline
		\rule[14pt]{0pt}{0pt}
%		$\delta_{CP}^\ell$ &  $328.932^\circ$& $ 170.523^\circ$ & $335.678^\circ$\\
		$[\delta_{CP}^\ell,\ \alpha_{21},\,\alpha_{31}]$ &  $[272^\circ,\, 222^\circ,\, 120^\circ]$    \\ \hline
		\rule[14pt]{0pt}{0pt}
%		$[\alpha_{21},\,\alpha_{31}]$ &  $[169.96^\circ,\, 336.337^\circ]$ & $[ 167.63^\circ,\, 159.805^\circ]$ &  $[ 157.025^\circ,\, 130.115^\circ]$	\\	
%		$[\alpha_{21},\,\alpha_{31}]$ &  $[0.878^\circ,\, 352^\circ]$   \\	 \hline		\rule[14pt]{0pt}{0pt}
%		$\sum m_i$ &  $71.3811$\,meV &	 $105.101$\,meV & $113.717$\, meV \\
		$\sum m_i$ &  $64.6$\,meV    \\ \hline
		\rule[14pt]{0pt}{0pt}
		$\langle m_{ee} \rangle$ &  $3.28$\,meV    \\ \hline
		\rule[14pt]{0pt}{0pt}
		$\sqrt{\Delta\chi^2}$ &  $2.05$   \\ \hline
		\hline
	\end{tabular}
	\caption{Numerical benchmark point of our input parameters and observables in NH. Here, we take it such that $\sqrt{\Delta \chi^2}$ is {the minimum}.}
	\label{bp-tab_nh-2}
\end{table}
%%%%%%%%%%%%%%%%%%%%%%%%%%%%%%%%%%%%%%%%%%%%%%%%%%%%%%%
%
Finally, we show a benchmark point for NH in Table~\ref{bp-tab_nh-2}, where the benchmark is taken such that $\sqrt{\Delta \chi^2}$ is minimum. %On the other hand, NH2 is taken so that $\delta_{CP}$ is closest to the best fit (BF) value of $195^\circ$ within $\sqrt{\Delta \chi^2}\le2$.

\subsubsection{$IH$ \label{sec:NA1}}
%

%------------------------------------------------------------------
\begin{figure}[htbp]
  \includegraphics[width=77mm]{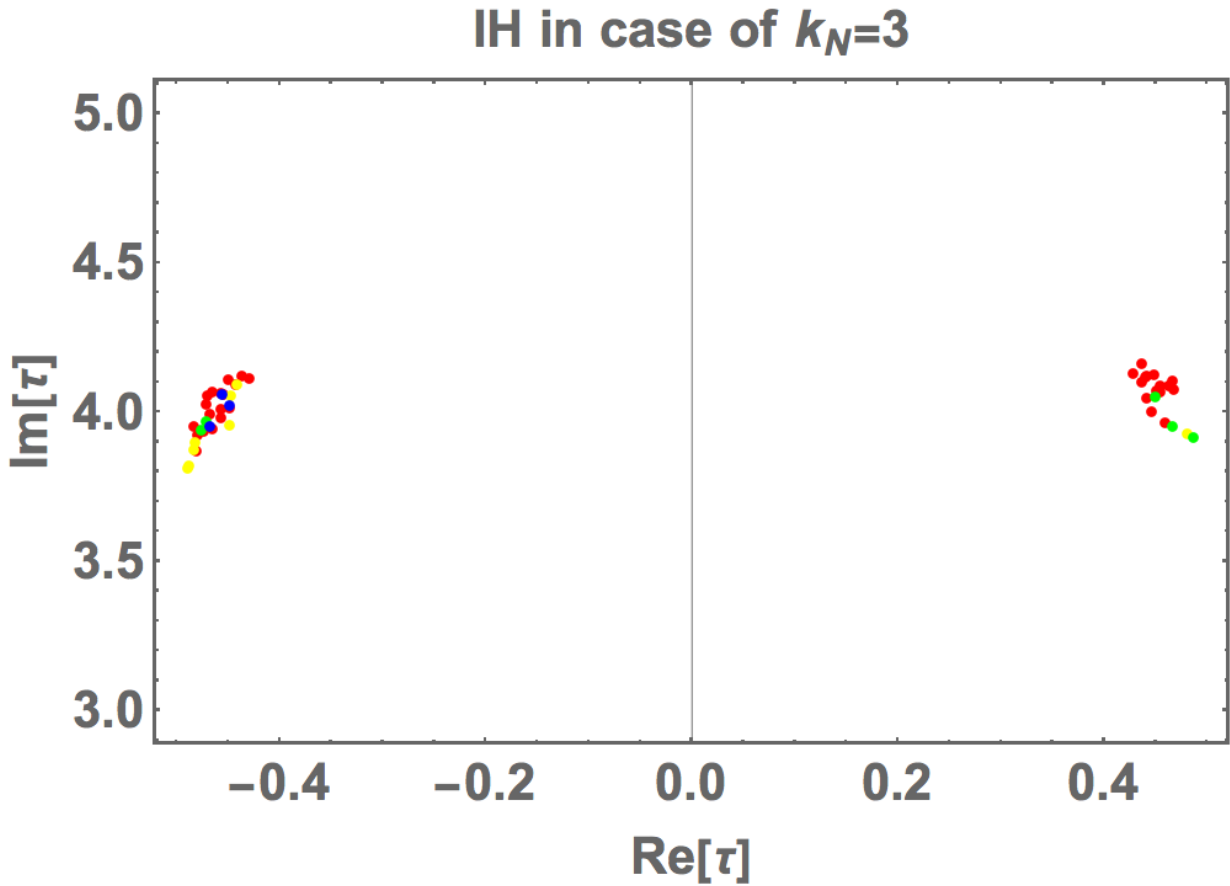}
  \caption{The scatter plots for the real $\tau$ and imaginary $\tau$ at nearby the fixed point $\tau=\omega$ in IH.
    In the $\Delta\chi^2$ analysis,  yellow color is for $2-3$, and red one is for $3-5$ of $\sqrt{\Delta\chi^2}$. The black solid line is the boundary of the fundamental domain at $|\tau|=1$.}
  \label{fig:tau_ih-2}
\end{figure}
%%%
In Fig.~\ref{fig:tau_ih-2}, we show the scatter plots of the real $\tau$ and imaginary $\tau$ in {the case} of IH and $k_N=3$.
The solid line is the fundamental domain boundary at $|\tau|=1$. Here, the allowed region is localized at {nearby} $\pm\frac12 +4 i$.

\begin{figure}[htbp]
  \includegraphics[width=53.5mm]{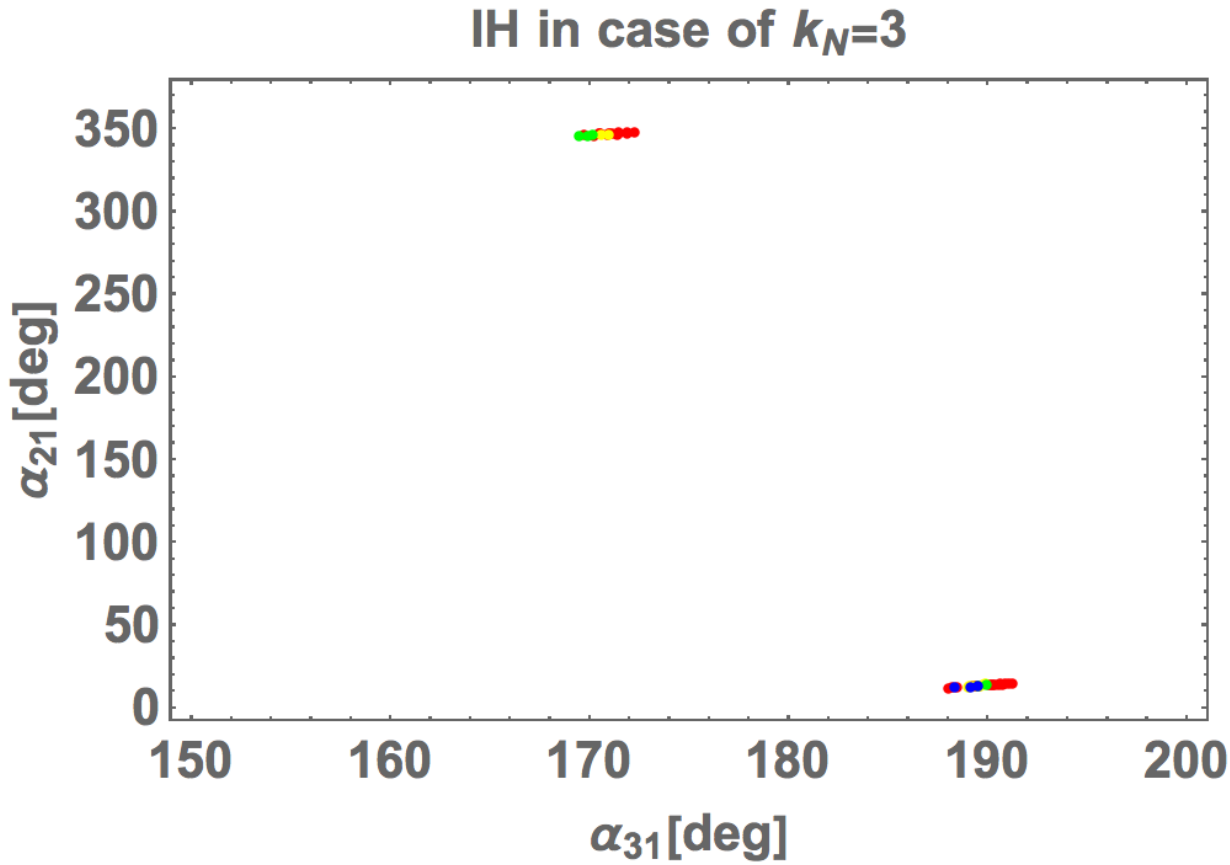}
  \includegraphics[width=53.5mm]{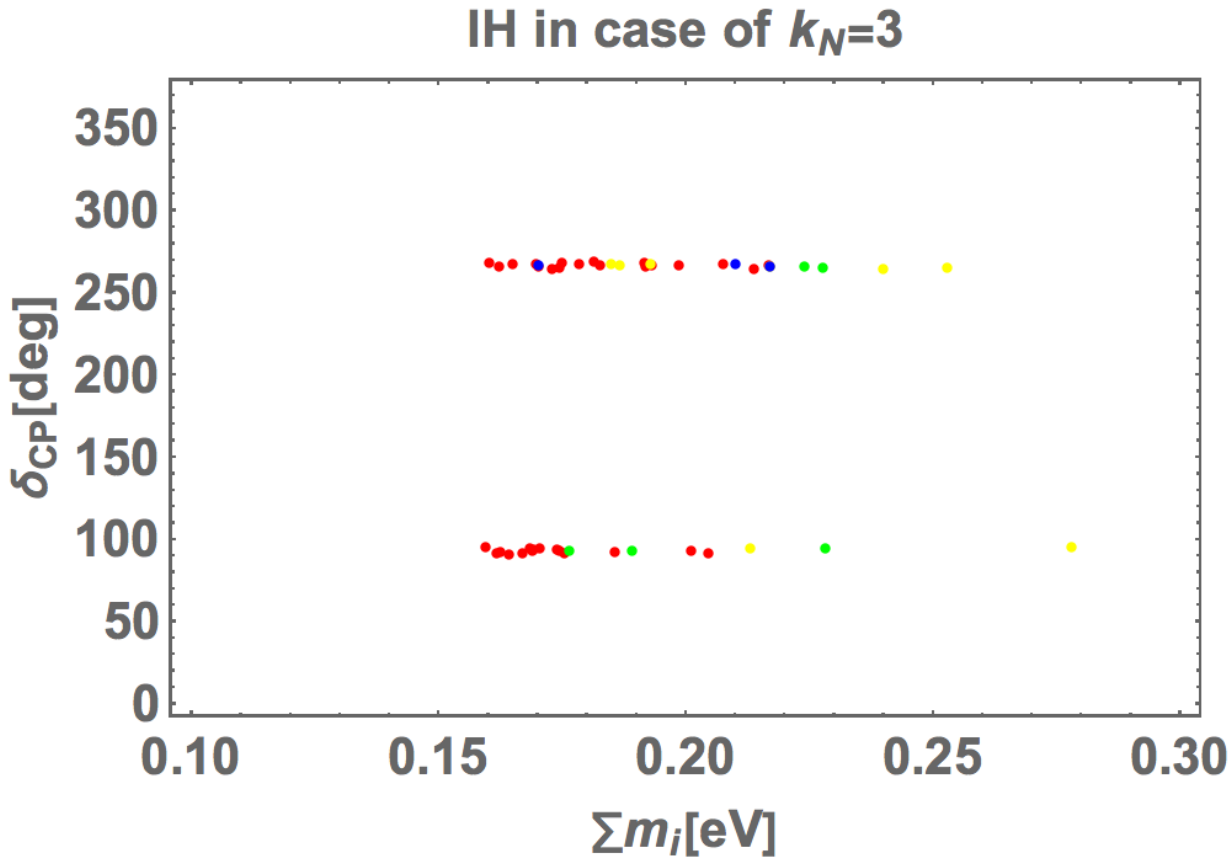}
  \includegraphics[width=53.5mm]{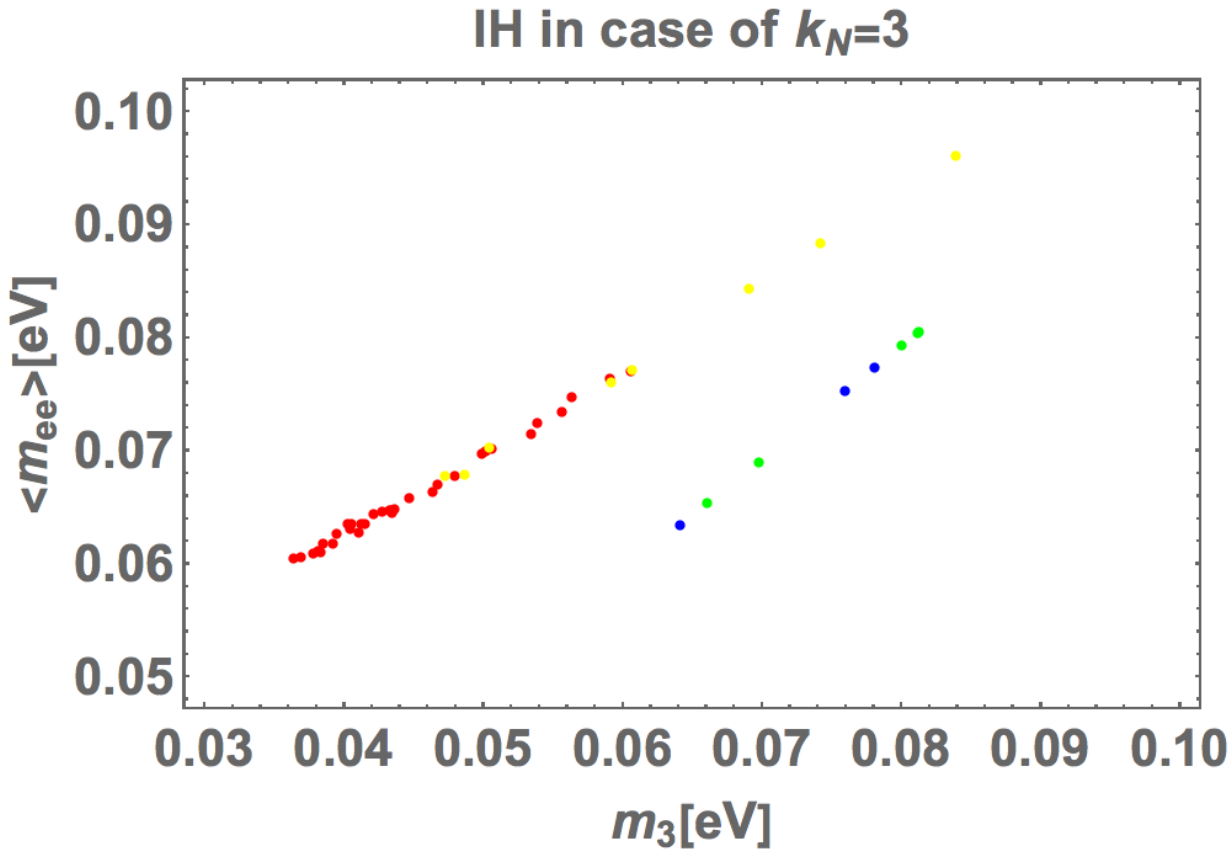}
  \caption{The scatter plots for the Majorana phases $\alpha_{31}$ and  $\alpha_{21}$ (left panel), the summation of active neutrino mass eigenstates and Dirac CP phase (middle panel), and the lightest mass of active neutrinos ($m_3$) and neutrinoless double beta decay ($\langle m_{ee}\rangle$) (right panel) in case of IH. The color representations are the same as in Fig.~\ref{fig:tau_ih-1}.}
  \label{figs-obs_ih-2}
\end{figure}
%%%%%%
In Fig.~\ref{figs-obs_ih-2}, we show the scatter plots for the Majorana phases $\alpha_{31}$ and  $\alpha_{21}$ (left panel), the summation of active neutrino mass eigenstates and Dirac CP phase (middle panel), and the lightest mass of active neutrinos ($m_1$) and neutrinoless double beta decay ($\langle m_{ee}\rangle$) (right panel) in case of NH. The color representations are the same as in Fig.~\ref{fig:tau_nh-1}.
%%%
From the left panel, we find that there are two islands; $\alpha_{31}=[188-192]$ deg for $\alpha_{21}\simeq20$ deg
and $\alpha_{31}=[169-173,]$ deg for $\alpha_{21}\simeq350$ deg.
%%%
From the middle panel, we also find two archipelagos; $\delta_{CP}=[100, 275]$ deg for $\sum_{i}m_i =[0.16-0.28]$ eV. 
This implies that all the regions would be excluded by bound on cosmological constant; $\sum m_i\le120$ meV.
%%%
From the right panel, we find that $m_3=[35-85]$ meV and $\langle m_{ee}\rangle=[60-95]$ meV.
%%%

%%%%%%
\begin{figure}[htbp]
  \includegraphics[width=77mm]{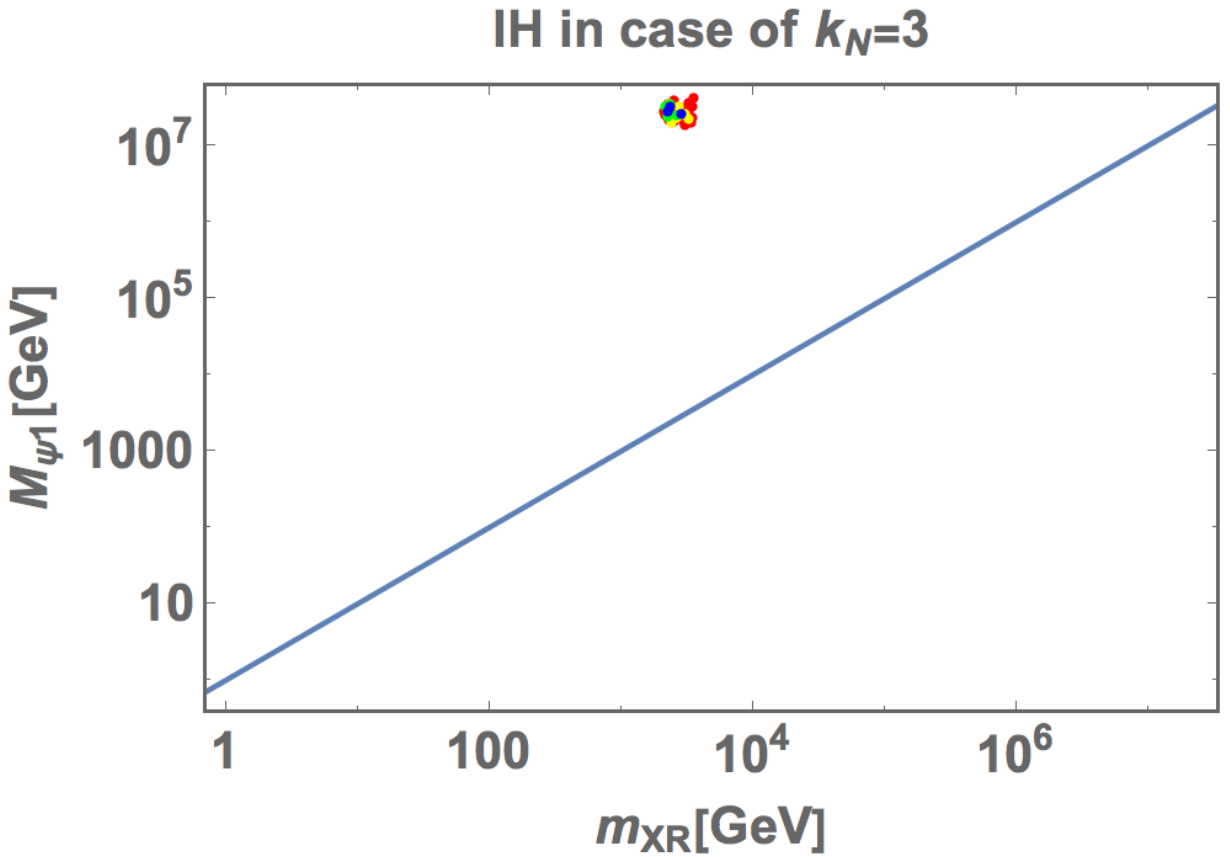}
   \caption{The scatter plots for $m_{\chi_R}$ and $M_{\psi_1}$. The color representations are the same as in Fig.~\ref{fig:tau_nh-1} and the blue line shows $m_{\chi_R}=M_{\psi_1}$.}
  \label{figs-mdm_ih-2}
\end{figure}
In Fig.~\ref{figs-mdm_ih-2}, we show the scattering plots in {the} case of IH that is {the same} as the case of NH. The color representations are the same as in Fig.~\ref{fig:tau_ih-2}.
%%%
In Fig.~\ref{figs-mdm_ih-2}, we show the scatter plots for the DM candidates $M_1$ and $m_{\chi_R}$ in {the case} of NH. 
Here, the color representations are the same as in Fig.~\ref{fig:tau_nh-1}.
This figure represents that DM $\chi_{R}$ works well. 
%%%
{
This tight relation might come from the small allowed space on tau. 
Moreover, we found the three mass eigenvalues of $\psi$ are almost degenerate, which suggests the $U_{\rm{PMNS}}$ mixings are determined by the Yukawa coupling $\tilde F$ only and the loop function becomes the overall factor. Since the overall factor is used to fit the observable atmospheric neutrino mass-squared difference, the small region would reflect the error bar of this experimental value up to $3\sigma$ interval.
}

%%%%%%%%%%%%%%%%%%%%%%%%%%%%%%%%%%%%%%%%%%%%%%%%%%%
\begin{table}[h]
	\centering
		\begin{tabular}{|c|c|} \hline 
			\rule[14pt]{0pt}{0pt}
%Lepton		&  NH($\tau\approx 1.75 i$) & IH($\tau\approx i$)  & IH($\tau\approx 1.76 i$)\\  \hline
 		&  IH  \\  \hline
			\rule[14pt]{0pt}{0pt}
		$\tau$ & $-0.456 + 4.06 i$      \\ \hline
		\rule[14pt]{0pt}{0pt}
		%%%
%		
		$[b_\eta/a_\eta,c_\eta/a_\eta]$ & $[0.105,\ -18.7]$      \\ \hline
		\rule[14pt]{0pt}{0pt}
		$[ r_1, r_2, \epsilon]$ & $[-0.00156 - 0.511 i,\ -0.00146 - 51.6 i ,\ 169 - 0.00202 i]$      \\ \hline
		\rule[14pt]{0pt}{0pt}
%		$c_\eta$  &  $0.00767509 - 0.0313634 i$ & $0.0013744 - 0.0000470787 I$ & $-0.00227049 + 0.00350111 I$\\
%		$\epsilon$ & $-0.0619 + 0.00436  i$ & $-0.0626 - 0.00268 i$     \\ \hline
%		\rule[14pt]{0pt}{0pt}
		%%%
%		$a$  &  $0.000156626 + 0.000114523 i$ & $3.04135 + 0.98336 I$ & $0.00853417 + 0.0246743 I$\\
		$M_0/{\rm GeV}$ & $1.61\times10^6$    \\ \hline
		\rule[14pt]{0pt}{0pt}
%		$b$  &  $-0.0173463 + 0.0012792 i$ & $0.000143616 + 0.000153805 I$ & $-0.0960243 + 0.0408818 I$\\
		$[M_1,M_2,M_3]/ {\rm GeV}$  &  $[2.77\times10^7, 2.78\times10^7, 2.78\times10^7] $     \\ \hline
		\rule[14pt]{0pt}{0pt}
		%$b$  &  $-0.0173463 + 0.0012792 i$ & $0.000143616 + 0.000153805 I$ & $-0.0960243 + 0.0408818 I$\\
		$[m_{\chi_R},m_{\chi_I},m_{\eta_1}]/ {\rm GeV}$  &  $[2260,\ 2980,\ 52700] $    \\ \hline
		\rule[14pt]{0pt}{0pt}
%		$c$  &  $-0.000300709 - 0.000100567 i$ & $0.005563 - 0.0021227 I$ & $-1.78167 + 4.54256 I$\\
		$\kappa/ {\rm GeV}$  &  $4.92\times10^{14} $   \\ \hline
		\rule[14pt]{0pt}{0pt}
%		$b'$  &  $0.000137596 - 0.0000163875 i$ & $0.0000236483 + 0.0000158969 I$ & $-0.000105682 + 0.000106792 
		$\Delta m^2_{\rm atm}$  &  $2.47\times10^{-3} {\rm eV}^2$   \\ \hline
		\rule[14pt]{0pt}{0pt}
%		$c'$  &  $-0.213667 - 0.271705 i$ & $-0.153721 + 0.0313641 I$ & $-3.40695 \times 10^{-6} - 0.0000598624 I$\\
		$\Delta m^2_{\rm sol}$  &  $7.58\times10^{-5} {\rm eV}^2$        \\ \hline
		\rule[14pt]{0pt}{0pt}
%		$\sin^2\theta_{12}$ & $ 0.322231$& $0.28036$ & $0.33157$\\
		$\sin\theta_{12}$ & $ 0.568$   \\ \hline
		\rule[14pt]{0pt}{0pt}
%		$\sin^2\theta_{23}$ &  $ 0.563489$& $0.463487$ & $0.579596$\\
		$\sin\theta_{23}$ &  $ 0.776$     \\ \hline
		\rule[14pt]{0pt}{0pt}
%		$\sin^2\theta_{13}$ &  $ 0.0235092$&$0.0240532$ & $0.0218055$\\
		$\sin\theta_{13}$ &  $ 0.150$    \\ \hline
		\rule[14pt]{0pt}{0pt}
%		$\delta_{CP}^\ell$ &  $328.932^\circ$& $ 170.523^\circ$ & $335.678^\circ$\\
		$[\delta_{CP}^\ell,\ \alpha_{21},\,\alpha_{31}]$ &  $[267^\circ,\, 12.1^\circ,\, 188^\circ]$    \\ \hline
		\rule[14pt]{0pt}{0pt}
%		$[\alpha_{21},\,\alpha_{31}]$ &  $[169.96^\circ,\, 336.337^\circ]$ & $[ 167.63^\circ,\, 159.805^\circ]$ &  $[ 157.025^\circ,\, 130.115^\circ]$	\\	
%		$[\alpha_{21},\,\alpha_{31}]$ &  $[0.878^\circ,\, 352^\circ]$   \\	 \hline		\rule[14pt]{0pt}{0pt}
%		$\sum m_i$ &  $71.3811$\,meV &	 $105.101$\,meV & $113.717$\, meV \\
		$\sum m_i$ &  $170$\,meV    \\ \hline
		\rule[14pt]{0pt}{0pt}
		$\langle m_{ee} \rangle$ &  $63.4$\,meV    \\ \hline
		\rule[14pt]{0pt}{0pt}
		$\sqrt{\Delta\chi^2}$ &  $1.75$   \\ \hline
		\hline
	\end{tabular}
	\caption{Numerical benchmark point of our input parameters and observables in IH. Here, we take it such that $\sqrt{\Delta \chi^2}$ is the minimum.}
	\label{bp-tab_ih-2}
\end{table}
%%%%%%%%%%%%%%%%%%%%%%%%%%%%%%%%%%%%%%%%%%%%%%%%%%%%%%%
%
%%%%%%%%%%%%%%%%%%%%%%%%%%%%%%%%%%%%%%%%%%%%%%%%%%%%%%%
%
Finally, we show a benchmark point for IH in Table~\ref{bp-tab_ih-2}, where the benchmark is taken such that $\sqrt{\Delta \chi^2}$ is minimum. %On the other hand, IH2 is taken so that $\delta_{CP}$ is closest to the BF value of $286^\circ$ within $\sqrt{\Delta \chi^2}\le6$.

}

%%%%%%%%%%%%%%%%%%%%%%%%%%%%%%%%%%%%%%%%%%%%%%%%%%
\subsection{Numerical analysis in muon $g-2$ and dark matter \label{sec:NA2}}
%%%%%%%%%%%%%%%%%%%%%%%%%%%%%%%%%%%%%%%%%%%%%%%%%%
In this subsection, we present our numerical analysis satisfying the muon $g-2$ and relic density of $\chi_R$,
where we have fixed the other parameters in Tab.~\ref{bp-tab_nh-2} and $A_b v_1/\sqrt2=m_\chi^2/10$ GeV$^2$.
%=500\ {\rm GeV}\times 245$ GeV.
Here, we do not apply IH since it does not satisfy the cosmological constant.
% and Tab.~\ref{bp-tab_ih-2} for IH.
%
As we mentioned before, the muon $g-2$ is within the range of 3$\sigma$ interval in Eq.~(\ref{eq:yeg2}).
While the correct relic density requires the thermally averaged {cross-section} to be the range of $[1.5-3.5]\times 10^{-9}$ GeV$^{-2}$.
Here we take the following ranges of the input parameters;
\begin{align}
&\{h ,\ y_E \} \in [10^{-5},1],\    M_E \in [200,500]{\rm GeV}, 
\end{align}
\begin{figure}[htbp]
  \includegraphics[width=53mm]{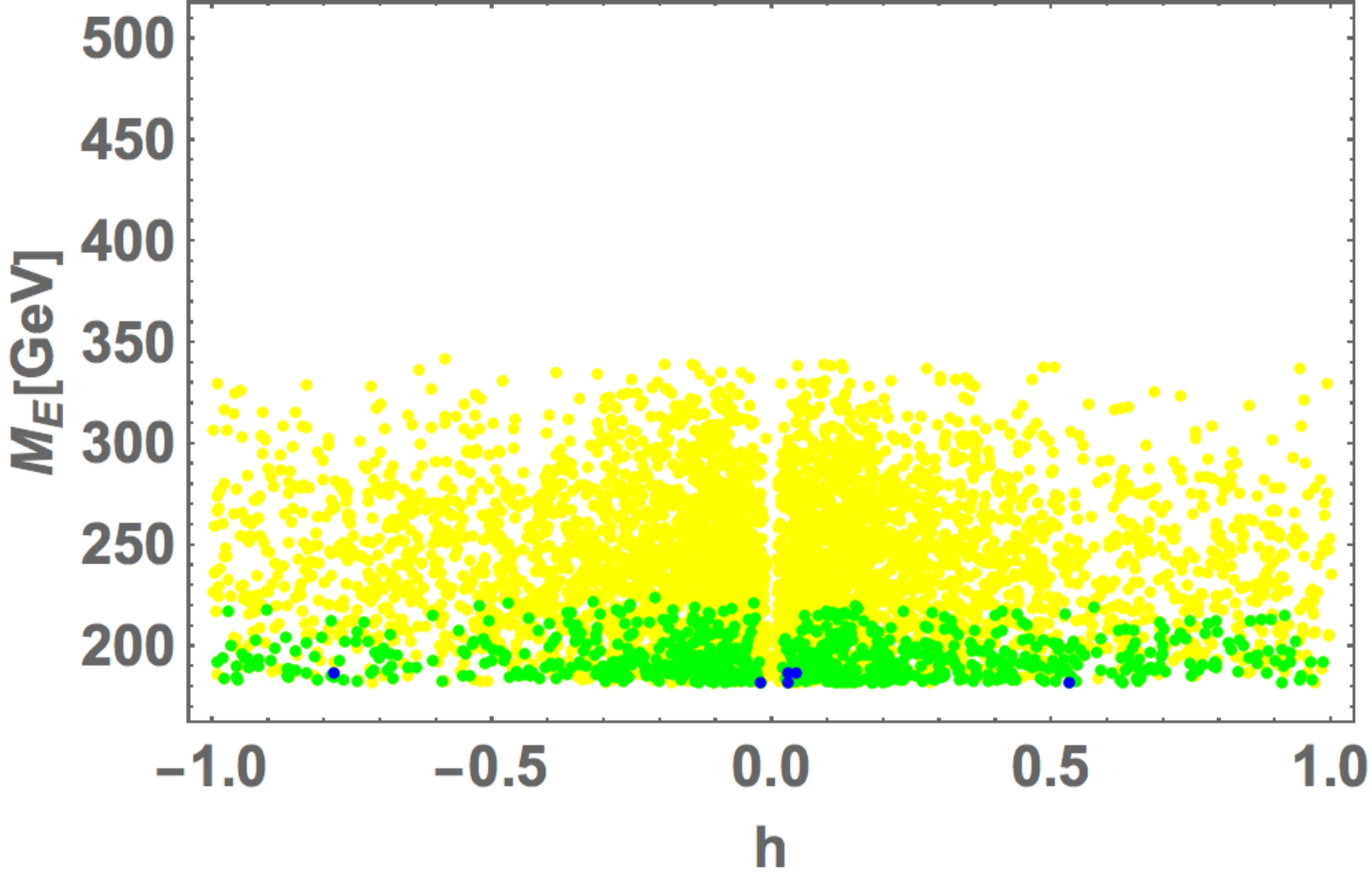}
  \includegraphics[width=53mm]{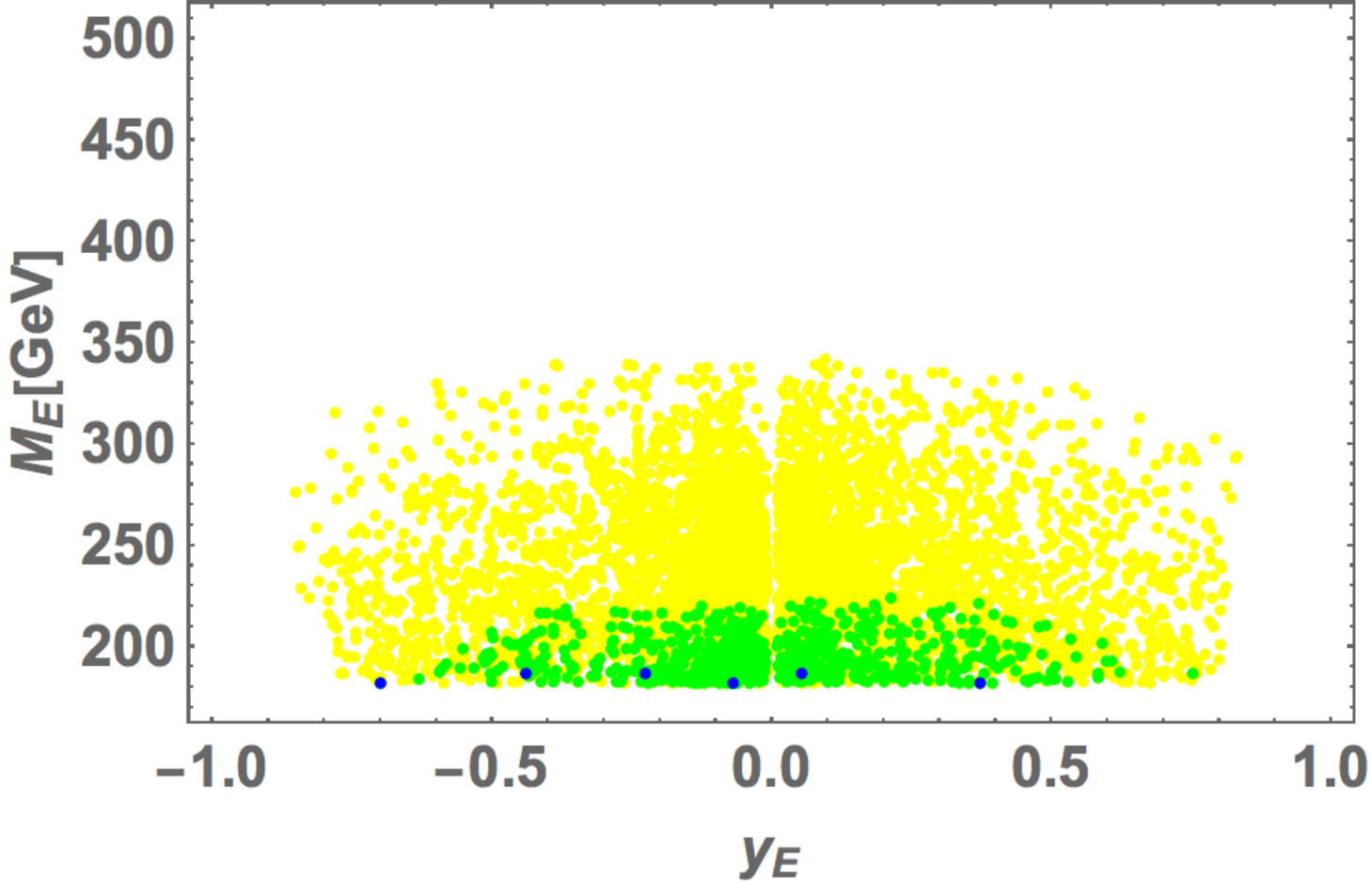}
  \includegraphics[width=53mm]{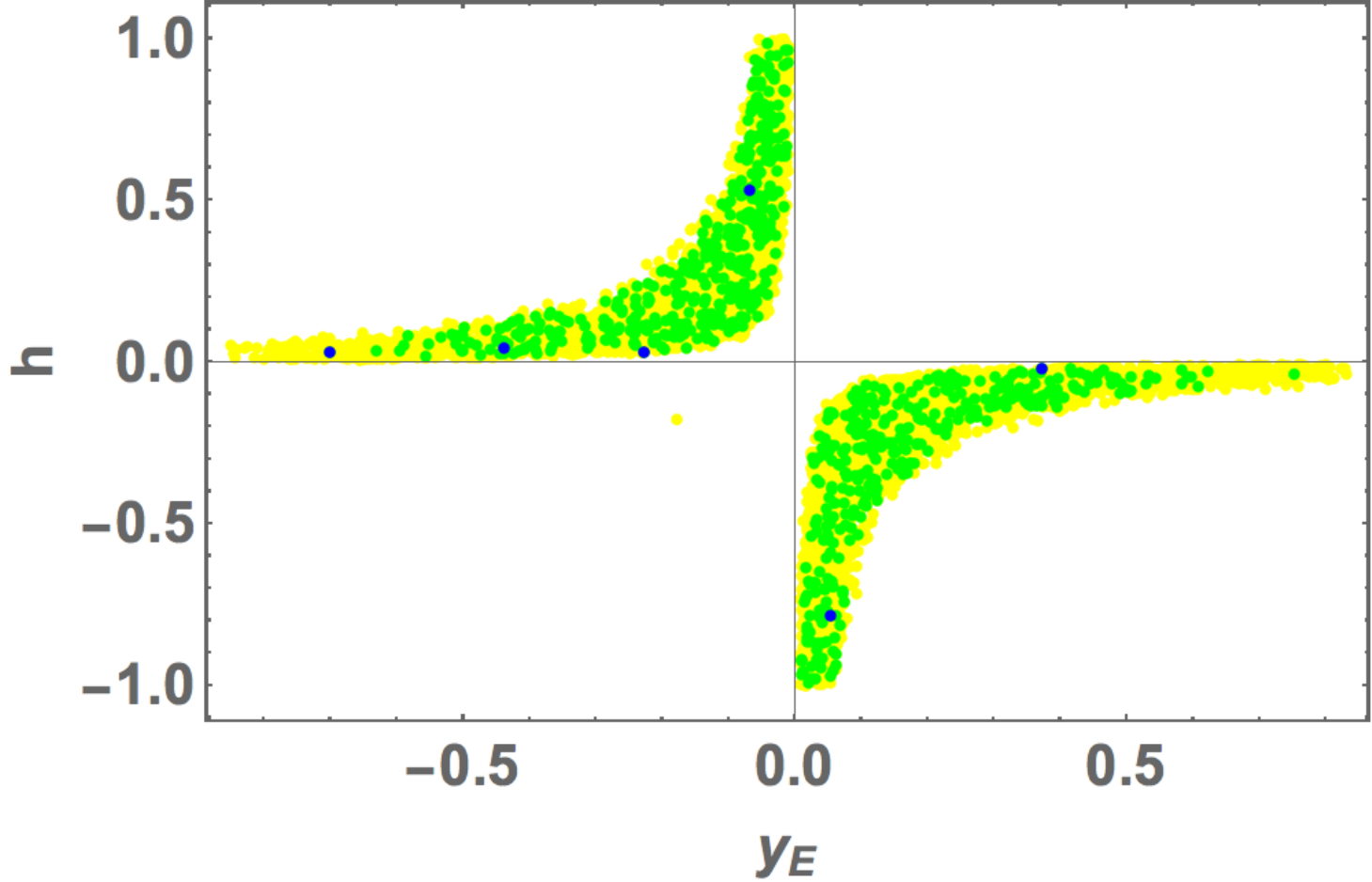}
  \caption{The scatter plots satisfy the DM relic density and the muon $g-2$. The left panel shows the allowed space of $h$ and $M_E$ [GeV], the middle one $y_E$ and $M_E$ [GeV],  and the right one  $y_E$ and $h$. The blue, yellow, and red colors respectively represent 1, 2, and 3$\sigma$ interval in Eq.(\ref{eq:yeg2}).}
  \label{figs.mx-damu}
\end{figure}
%%%%%%
In Fig.~\ref{figs.mx-damu}, we show the allowed space of $h$ and $M_E$ GeV (left panel),
$y_E$ and $M_E$ GeV (middle panel), and $y_E$ and $h$ (right panel),
where the blue, yellow, and red colors respectively represent1, 2, and 3$\sigma$ interval of muon $g-2$ in Eq.(\ref{eq:yeg2}).
These figures suggest that $h$ and $y_E$ can be appropriate values, but $M_E$ has to be rather small that is $200$ GeV$\lesssim M_E\lesssim$340 GeV.
%%%%%%%%%%%%%%%%%%%%%%%%
{
The mass bound on $E$ could be estimated by the analogy of sleptons at the Large Hadron Collider in ref.~\cite{CMS:2020bfa}. Our $E$ mainly decays into a muon and missing particle $\chi_R$.
Following the figure 14 in ref.~\cite{CMS:2020bfa},
our mass range of $E$ may still be allowed when the mass of $\chi_R$ is enough heavy; ${\rm e.g.}$, 100 GeV$\le m_{\chi_R}$ for $M_E\simeq200$ GeV. This condition satisfies in Tab.~\ref{bp-tab_nh-2},
although the detailed analysis is beyond our scope.
}
%%%%%%%%%%%%%%%%%%%%%%%%

\section{Conclusions and discussions} \label{con}
%%%

We investigated the successful model to explain the muon anomalous magnetic moment from Yukawa contributions. 
Thanks to the modular $A_4$ symmetry, any lepton flavor violation processes that spoil our model are forbidden. 
We have studied a predictive radiative seesaw model including a dark matter candidate.
At first, we have constructed the minimum model ($k_N=1$) to satisfy the neutrino oscillation data and obtain several predictions such as Dirac CP and Majorana phases, the neutrino masses through $\chi^2$ analysis.
However, the model with $k_N=1$  would not have provided a proper DM candidate.
Therefore, we have minimally extended the model ($k_N=3$) and got the correct relic density of DM.
In the NH case of $k_N=3$ framework, we have shown the allowed regions to satisfy the muon anomalous magnetic moment and the observed relic density in addition to predictions of the lepton sector, where IH does not satisfy the constraint on the cosmological constant.

%%%%%%%%%%%%%%%%%%%%%%%%%%%%%%%%%%%
%\vspace{0.5cm}
%\hspace{0.2cm} 

\begin{acknowledgments}
The work of P.T.P.H was supported by the National Research Foundation of Korea (NRF)
grant funded by the Korea government (MSIT) No.~2018R1A5A1025563 and No.~2019R1A2C1005697.
The work of J.K. is supported in part by the Korean Institute Advanced Studies (KIAS) Individual Grant. No. PG074201.
The work of D.W.K. is supported in part by the KIAS Individual Grant. No. PG076201.
The work of H.O. was supported by the Junior Research Group (JRG) Program at the Asia-Pacific Center for Theoretical
Physics (APCTP) through the Science and Technology Promotion Fund and Lottery Fund of the Korean Government and was supported by the Korean Local Governments-Gyeongsangbuk-do Province and Pohang City.
H.O. is sincerely grateful {to} all the KIAS members.
\end{acknowledgments}

\section*{Appendix}

 %%%%%%%%%%%%%%%%%%%%%%%%%%%%%%%%%%%%%%%%%%%%%%%%%%%%%%%%%%%
In this appendix, we present several properties of the modular $A_4$ symmetry. 
In general, the modular group $\bar\Gamma$ is a group of the linear fractional transformation
$\gamma$, acting on the modulus $\tau$ 
which belongs to the upper-half complex plane and transforms as
\begin{equation}\label{eq:tau-SL2Z}
\tau \longrightarrow \gamma\tau= \frac{a\tau + b}{c \tau + d}\ ,~~
{\rm where}~~ a,b,c,d \in \mathbb{Z}~~ {\rm and }~~ ad-bc=1, 
~~ {\rm Im} [\tau]>0 ~.
\end{equation}
This is isomorphic to  $PSL(2,\mathbb{Z})=SL(2,\mathbb{Z})/\{I,-I\}$ transformation.
Then modular transformation is generated by two transformations $S$ and $T$ defined by:
\begin{eqnarray}
S:\tau \longrightarrow -\frac{1}{\tau}\ , \qquad\qquad
T:\tau \longrightarrow \tau + 1\ ,
\end{eqnarray}
and they satisfy the following algebraic relations, 
\begin{equation}
S^2 =\mathbb{I}\ , \qquad (ST)^3 =\mathbb{I}\ .
\end{equation}
More concretely, we fix the basis of $S$ and $T$ as follows:
  \begin{align}
S=\frac13
 \begin{pmatrix}
 -1 & 2 & 2  \\
 -2 & -1 & 2  \\
 2 & 2 & -1  \\
 \end{pmatrix} ,\quad 
 T= 
 \begin{pmatrix}
 1 & 0 & 0 \\
0 & \omega & 0  \\
0 & 0 & \omega^2  \\
 \end{pmatrix} ,
 \end{align}
where $\omega\equiv e^{2\pi i/3}$.

Thus, we introduce the series of groups $\Gamma(N)~ (N=1,2,3,\dots)$ that is so-called "principal congruence subgroups of $SL(2,Z)$", which are defined by
 \begin{align}
 \begin{aligned}
 \Gamma(N)= \left \{ 
 \begin{pmatrix}
 a & b  \\
 c & d  
 \end{pmatrix} \in SL(2,\mathbb{Z})~ ,
 ~~
 \begin{pmatrix}
  a & b  \\
 c & d  
 \end{pmatrix} =
  \begin{pmatrix}
  1 & 0  \\
  0 & 1  
  \end{pmatrix} ~~({\rm mod}~N) \right \}
 \end{aligned},
 \end{align}
and we define $\bar\Gamma(2)\equiv \Gamma(2)/\{I,-I\}$ for $N=2$.
Since the element $-I$ does not belong to $\Gamma(N)$
  for $N>2$ case, we have $\bar\Gamma(N)= \Gamma(N)$,
  that {is an infinite} normal subgroup of $\bar \Gamma$ known as principal congruence subgroups.
   We thus obtain finite modular groups as the quotient groups defined by
   $\Gamma_N\equiv \bar \Gamma/\bar \Gamma(N)$.
For these finite groups $\Gamma_N$, $T^N=\mathbb{I}$  is imposed, and
the groups $\Gamma_N$ with $N=2,3,4$ and $5$ are isomorphic to
$S_3$, $A_4$, $S_4$ and $A_5$, respectively \cite{deAdelhartToorop:2011re}.

Modular forms of level $N$ are 
holomorphic functions $f(\tau)$ which are transformed under the action of $\Gamma(N)$ given by
\begin{equation}
f(\gamma\tau)= (c\tau+d)^k f(\tau)~, ~~ \gamma \in \Gamma(N)~ ,
\end{equation}
where $k$ is the so-called modular weight.

%Here we discuss the modular symmetric theory framework without imposing supersymmetry explicitly, considering the $A_4$ ($N=3$) modular group. 
Under the modular transformation in Eq.(\ref{eq:tau-SL2Z}) in case of $A_4$ ($N=3$) modular group, a field $\phi^{(I)}$ is also transformed as 
\begin{equation}
\phi^{(I)} \to (c\tau+d)^{-k_I}\rho^{(I)}(\gamma)\phi^{(I)},
\end{equation}
where  $-k_I$ is the modular weight and $\rho^{(I)}(\gamma)$ denotes a unitary representation matrix of $\gamma\in\Gamma(2)$ ($A_4$ representation).
Thus Lagrangian such as Yukawa terms can be invariant if the sum of modular weight from fields and modular form in {the corresponding} term is zero (also invariant under $A_4$ and gauge symmetry).

The kinetic terms and quadratic terms of scalar fields can be written by 
\begin{equation}
\sum_I\frac{|\partial_\mu\phi^{(I)}|^2}{(-i\tau+i\bar{\tau})^{k_I}} ~, \quad \sum_I\frac{|\phi^{(I)}|^2}{(-i\tau+i\bar{\tau})^{k_I}} ~,
\label{kinetic}
\end{equation}
which is invariant under the modular transformation and the overall factor is eventually absorbed by a field redefinition consistently.
Therefore the Lagrangian associated with these terms should be invariant under the modular symmetry.

The basis of modular forms with weight 2, $Y^{(2)}_3 = (y_{1},y_{2},y_{3})$, transforming
as a triplet of $A_4$ is written in terms of Dedekind eta-function  $\eta(\tau)$ and its derivative \cite{Feruglio:2017spp}:
%%%%%%%%%%%%%%%%%%%%%%%
\begin{align} 
\label{eq:Y-A4}
y_{1}(\tau) &= \frac{i}{2\pi}\left( \frac{\eta'(\tau/3)}{\eta(\tau/3)}  +\frac{\eta'((\tau +1)/3)}{\eta((\tau+1)/3)}  
+\frac{\eta'((\tau +2)/3)}{\eta((\tau+2)/3)} - \frac{27\eta'(3\tau)}{\eta(3\tau)}  \right)\nn\\ 
&\simeq
1+12 q+36 q^2+12 q^3+\cdots,\\
y_{2}(\tau) &= \frac{-i}{\pi}\left( \frac{\eta'(\tau/3)}{\eta(\tau/3)}  +\omega^2\frac{\eta'((\tau +1)/3)}{\eta((\tau+1)/3)}  
+\omega \frac{\eta'((\tau +2)/3)}{\eta((\tau+2)/3)}  \right) , \label{eq:Yi} \nn\\ 
&\simeq
-6q^{1/3} (1+7 q+8 q^2+\cdots),\\
y_{3}(\tau) &= \frac{-i}{\pi}\left( \frac{\eta'(\tau/3)}{\eta(\tau/3)}  +\omega\frac{\eta'((\tau +1)/3)}{\eta((\tau+1)/3)}  
+\omega^2 \frac{\eta'((\tau +2)/3)}{\eta((\tau+2)/3)}  \right)\nn\\ 
&\simeq
-18q^{2/3} (1+2 q+5 q^2+\cdots),
\end{align}
where $q=e^{2\pi i \tau}$, and expansion form in terms of $q$ would sometimes be useful to have numerical analysis.
%%%%%%%%%%%%%%%%%%%%%
% Notice here that any singlet couplings under $A_4$ start from $-k=4$ constructed from the modular forms with $-k=2$ while it is absent if $-k=2$.

Then, we can construct the higher order of couplings; e.g., $Y^{(4)}_3, Y^{(6)}_3, Y^{(6)}_{3'}$ following the multiplication rules as follows:
\begin{align}
%Y^{(4)}_1& = y^2_1+2 y_2 y_3, \ Y^{(6)}_1 = y^3_1+y^3_2 + y^3_3 -3 y_1y_2y_3, \ Y^{(10)}_1 = Y^{(4)}_1 Y^{(6)}_1,\\
%%%
Y^{(4)}_3&\equiv (y_1^{(4)},y_2^{(4)},y_3^{(4)}) 
= (y^2_1 - y_2 y_3, y^2_3 - y_1 y_2, y^2_2 - y_1 y_3),\\
Y^{(6)}_3&\equiv (y_1^{(6)},y_2^{(6)},y_3^{(6)}) = ( y^3_1+2y_1 y_2 y_3, y_1^2y_2+2 y^2_2 y_3, y^2_1 y_3+2 y^2_3 y_2),\\
Y^{(6)}_{3'}
&\equiv (y'^{(6)}_1,y'^{(6)}_2,y'^{(6)}_3) = ( y^3_3+2y_1 y_2 y_3, y^2_3 y_1+2 y^2_1 y_2, y^2_3 y_2+2 y^2_2 y_1),
%%%
\end{align}
%%%
where  the above relations are constructed by the multiplication rules under $A_4$ as shown below:
\begin{align}
\begin{pmatrix}
a_1\\
a_2\\
a_3
\end{pmatrix}_{\bf 3}
\otimes 
\begin{pmatrix}
b_1\\
b_2\\
b_3
\end{pmatrix}_{\bf 3'}
&=\left (a_1b_1+a_2b_3+a_3b_2\right )_{\bf 1} 
\oplus \left (a_3b_3+a_1b_2+a_2b_1\right )_{{\bf 1}'} \nonumber \\
& \oplus \left (a_2b_2+a_1b_3+a_3b_1\right )_{{\bf 1}''} \nonumber \\
&\oplus \frac13
\begin{pmatrix}
2a_1b_1-a_2b_3-a_3b_2 \\
2a_3b_3-a_1b_2-a_2b_1 \\
2a_2b_2-a_1b_3-a_3b_1
\end{pmatrix}_{{\bf 3}}
\oplus \frac12
\begin{pmatrix}
a_2b_3-a_3b_2 \\
a_1b_2-a_2b_1 \\
a_3b_1-a_1b_3
\end{pmatrix}_{{\bf 3'}\  } \ , \nonumber \\
\nonumber \\
{\bf 1} \otimes {\bf 1} = {\bf 1} \ , \quad &
{\bf 1'} \otimes {\bf 1'} = {\bf 1''} \ , \quad
{\bf 1''} \otimes {\bf 1''} = {\bf 1'} \ , \quad
{\bf 1'} \otimes {\bf 1''} = {\bf 1} \ .
\end{align}

% Ref Style
% Including title
%\bibliographystyle{utphys}
%\bibliography{MA4_emug2.bib}
%=====================================================

%=====================================================
\end{document}